\documentclass[aps,prc,reprint,superscriptaddress,floatfix,
  showpacs]{revtex4-1}
\pdfoutput=1
\usepackage{slashed}
\usepackage{graphicx}
\usepackage{subfigure}
\usepackage{hyperref}

\begin{document}

\title{Near Threshold Neutral Pion Electroproduction at High Momentum Transfers and
Generalized Form Factors}

\newcommand*{\ANL}{Argonne National Laboratory, Argonne, Illinois 60439}
\newcommand*{\ANLindex}{1}
\affiliation{\ANL}
\newcommand*{\ASU}{Arizona State University, Tempe, Arizona 85287-1504}
\newcommand*{\ASUindex}{2}
\affiliation{\ASU}
\newcommand*{\CSUDH}{California State University, Dominguez Hills, Carson, CA 90747}
\newcommand*{\CSUDHindex}{3}
\affiliation{\CSUDH}
\newcommand*{\CANISIUS}{Canisius College, Buffalo, NY}
\newcommand*{\CANISIUSindex}{4}
\affiliation{\CANISIUS}
\newcommand*{\CMU}{Carnegie Mellon University, Pittsburgh, Pennsylvania 15213}
\newcommand*{\CMUindex}{5}
\affiliation{\CMU}
\newcommand*{\CUA}{Catholic University of America, Washington, D.C. 20064}
\newcommand*{\CUAindex}{6}
\affiliation{\CUA}
\newcommand*{\SACLAY}{CEA, Centre de Saclay, Irfu/Service de Physique Nucl\'eaire, 91191 Gif-sur-Yvette, France}
\newcommand*{\SACLAYindex}{7}
\affiliation{\SACLAY}
\newcommand*{\CNU}{Christopher Newport University, Newport News, Virginia 23606}
\newcommand*{\CNUindex}{8}
\affiliation{\CNU}
\newcommand*{\UCONN}{University of Connecticut, Storrs, Connecticut 06269}
\newcommand*{\UCONNindex}{9}
\affiliation{\UCONN}
\newcommand*{\EDINBURGH}{Edinburgh University, Edinburgh EH9 3JZ, United Kingdom}
\newcommand*{\EDINBURGHindex}{10}
\affiliation{\EDINBURGH}
\newcommand*{\FU}{Fairfield University, Fairfield CT 06824}
\newcommand*{\FUindex}{11}
\affiliation{\FU}
\newcommand*{\FIU}{Florida International University, Miami, Florida 33199}
\newcommand*{\FIUindex}{12}
\affiliation{\FIU}
\newcommand*{\FSU}{Florida State University, Tallahassee, Florida 32306}
\newcommand*{\FSUindex}{13}
\affiliation{\FSU}
\newcommand*{\Genova}{Universit\`a di Genova, 16146 Genova, Italy}
\newcommand*{\Genovaindex}{14}
\affiliation{\Genova}
\newcommand*{\GWUI}{The George Washington University, Washington, DC 20052}
\newcommand*{\GWUIindex}{15}
\affiliation{\GWUI}
\newcommand*{\ISU}{Idaho State University, Pocatello, Idaho 83209}
\newcommand*{\ISUindex}{16}
\affiliation{\ISU}
\newcommand*{\INFNFE}{INFN, Sezione di Ferrara, 44100 Ferrara, Italy}
\newcommand*{\INFNFEindex}{17}
\affiliation{\INFNFE}
\newcommand*{\INFNFR}{INFN, Laboratori Nazionali di Frascati, 00044 Frascati, Italy}
\newcommand*{\INFNFRindex}{18}
\affiliation{\INFNFR}
\newcommand*{\INFNGE}{INFN, Sezione di Genova, 16146 Genova, Italy}
\newcommand*{\INFNGEindex}{19}
\affiliation{\INFNGE}
\newcommand*{\INFNRO}{INFN, Sezione di Roma Tor Vergata, 00133 Rome, Italy}
\newcommand*{\INFNROindex}{20}
\affiliation{\INFNRO}
\newcommand*{\ORSAY}{Institut de Physique Nucl\'eaire ORSAY, Orsay, France}
\newcommand*{\ORSAYindex}{21}
\affiliation{\ORSAY}
\newcommand*{\ITEP}{Institute of Theoretical and Experimental Physics, Moscow, 117259, Russia}
\newcommand*{\ITEPindex}{22}
\affiliation{\ITEP}
\newcommand*{\JMU}{James Madison University, Harrisonburg, Virginia 22807}
\newcommand*{\JMUindex}{23}
\affiliation{\JMU}
\newcommand*{\KNU}{Kyungpook National University, Daegu 702-701, Republic of Korea}
\newcommand*{\KNUindex}{24}
\affiliation{\KNU}
\newcommand*{\LPSC}{LPSC, Universit\'e Joseph Fourier, CNRS/IN2P3, INPG, Grenoble, France}
\newcommand*{\LPSCindex}{25}
\affiliation{\LPSC}
\newcommand*{\UNH}{University of New Hampshire, Durham, New Hampshire 03824-3568}
\newcommand*{\UNHindex}{26}
\affiliation{\UNH}
\newcommand*{\NSU}{Norfolk State University, Norfolk, Virginia 23504}
\newcommand*{\NSUindex}{27}
\affiliation{\NSU}
\newcommand*{\OHIOU}{Ohio University, Athens, Ohio  45701}
\newcommand*{\OHIOUindex}{28}
\affiliation{\OHIOU}
\newcommand*{\ODU}{Old Dominion University, Norfolk, Virginia 23529}
\newcommand*{\ODUindex}{29}
\affiliation{\ODU}
\newcommand*{\RPI}{Rensselaer Polytechnic Institute, Troy, New York 12180-3590}
\newcommand*{\RPIindex}{30}
\affiliation{\RPI}
\newcommand*{\URICH}{University of Richmond, Richmond, Virginia 23173}
\newcommand*{\URICHindex}{31}
\affiliation{\URICH}
\newcommand*{\ROMAII}{Universit\`a di Roma Tor Vergata, 00133 Rome Italy}
\newcommand*{\ROMAIIindex}{32}
\affiliation{\ROMAII}
\newcommand*{\MSU}{Skobeltsyn Nuclear Physics Institute, 119899 Moscow, Russia}
\newcommand*{\MSUindex}{33}
\affiliation{\MSU}
\newcommand*{\SCAROLINA}{University of South Carolina, Columbia, South Carolina 29208}
\newcommand*{\SCAROLINAindex}{34}
\affiliation{\SCAROLINA}
\newcommand*{\JLAB}{Thomas Jefferson National Accelerator Facility, Newport News, Virginia 23606}
\newcommand*{\JLABindex}{35}
\affiliation{\JLAB}
\newcommand*{\UTFSM}{Universidad T\'{e}cnica Federico Santa Mar\'{i}a, Casilla 110-V Valpara\'{i}so, Chile}
\newcommand*{\UTFSMindex}{36}
\affiliation{\UTFSM}
\newcommand*{\GLASGOW}{University of Glasgow, Glasgow G12 8QQ, United Kingdom}
\newcommand*{\GLASGOWindex}{37}
\affiliation{\GLASGOW}
\newcommand*{\VIRGINIA}{University of Virginia, Charlottesville, Virginia 22901}
\newcommand*{\VIRGINIAindex}{38}
\affiliation{\VIRGINIA}
\newcommand*{\WM}{College of William and Mary, Williamsburg, Virginia 23187-8795}
\newcommand*{\WMindex}{39}
\affiliation{\WM}
\newcommand*{\YEREVAN}{Yerevan Physics Institute, 375036 Yerevan, Armenia}
\newcommand*{\YEREVANindex}{40}
\affiliation{\YEREVAN}
\newcommand*{\NOWCNU}{Christopher Newport University, Newport News, Virginia 23606}
\newcommand*{\NOWMSU}{Skobeltsyn Nuclear Physics Institute, 119899 Moscow, Russia}
\newcommand*{\NOWORSAY}{Institut de Physique Nucl\'eaire ORSAY, Orsay, France}
\newcommand*{\NOWINFNGE}{INFN, Sezione di Genova, 16146 Genova, Italy}
\newcommand*{\NOWROMAII}{Universit\`a di Roma Tor Vergata, 00133 Rome Italy}
\author{P.~Khetarpal}
\affiliation{\RPI}
\affiliation{\FIU}
\author{P.~Stoler}
\affiliation{\RPI}
\author{I.G.~Aznauryan}
\affiliation{\JLAB}
\affiliation{\YEREVAN}
\author{V.~Kubarovsky}
\affiliation{\JLAB}
\affiliation{\RPI}
\author{K.P.~Adhikari} 
\affiliation{\ODU}
\author{D.~Adikaram} 
\affiliation{\ODU}
\author{M.~Aghasyan} 
\affiliation{\INFNFR}
\author{M.J.~Amaryan} 
\affiliation{\ODU}
\author{M.D.~Anderson} 
\affiliation{\GLASGOW}
\author{S.~Anefalos~Pereira} 
\affiliation{\INFNFR}
% ...

\collaboration{The CLAS Collaboration}
\noaffiliation


\date{\today}

\begin{abstract}
  We report the measurement of near threshold neutral pion electroproduction
  cross sections and the extraction of the associated structure functions on 
  the proton in the kinematic range 
  $Q^2$ from $2$ to $4.5$ GeV$^2$ and $W$ from $1.08$ to $1.16$ GeV.
  These measurements allow us to access the dominant pion-nucleon $s$-wave
  multipoles $E_{0+}$ and $S_{0+}$ in the near-threshold region. In the 
  light-cone sum-rule framework (LCSR), these multipoles are related to the 
  generalized form factors $G_1^{\pi^0 p}(Q^2)$ and $G_2^{\pi^0 p}(Q^2)$.
  The data are compared to these generalized form factors and
  the results for $G_1^{\pi^0 p}(Q^2)$ are found to be in good agreement with 
  the LCSR predictions, but the level of agreement with $G_2^{\pi^0 p}(Q^2)$ 
  is poor. 
\end{abstract}


\maketitle


\section{Introduction\label{sec:introduction}}

Pion photo- and electroproduction on the nucleon $\gamma N \to \pi N$, 
$\gamma^{*}N \to \pi N$ close to threshold has been studied 
extensively since the 1950s both experimentally and theoretically. Exact 
predictions for the threshold cross sections and the axial form factor were 
pioneered by Kroll and Ruderman in 1954 for photo-production
and are known as the low energy theorem (LET) \cite{kr}. 
This LET provided
model independent predictions of cross sections for pion photoproduction
in the threshold region by applying gauge and Lorentz invariance 
\cite{drechtiat}. This was the first of the LET predictions to appear 
but was not without limitations.
This LET predictions were restricted only to charged pions and the 
$\pi^0$ contribution was shown to vanish in the `soft pion' limit, \emph{i.e.},
$m_{\pi} \sim p_\pi$. Here, $m_\pi$ and $p_\pi$ are the mass and momentum of
the pion. Additionally, these cross section predictions were
limited to diagrams with first order contributions in the pion-nucleon mass 
ratio.
In later years, using vanishing pion mass chiral symmetry ($m_\pi \to 0$), 
these predictions were extended to pion electroproduction for both charged and
neutral pions \cite{namlu,namsh}. 

Of course, a vanishing pion mass doesn't relate to the observed mass of the
pion (the pion to nucleon mass ratio $m_\pi/m_N \sim 1/7$), so 
higher order finite mass corrections to the LET 
were formulated in the late sixties and early seventies before the appearance 
of QCD. These also included contributions to the non-vanishing neutral pion 
amplitudes for the cross section.

In the late eighties and early nineties, experiments at 
Mainz \cite{mainz} obtained threshold pion photo-production data on 
$\gamma p \to \pi^{0} p$. The theoretical predictions of LETs at the time 
were inconsistent with the data at low photon energies.
With the emergence of chiral perturbation theory ($\chi$PT),
the scattering amplitudes and some physical observables were systematically
expanded in the low energy limit in powers of pion mass and momentum. Using 
this framework, the LET was re-derived to include contributions to the 
amplitudes 
from certain loop diagrams, which were lost when the expansion was performed 
in terms of the pion mass, as was done in the earlier works \cite{vz,schkoch}. 
Further electroproduction experiments 
at NIKHEF \cite{welch} on $\gamma^{*} p \to \pi^{0} p$ with photon virtuality 
$Q^2 \sim 0.05-0.1$ GeV$^2$ \footnote{For convenience, we use 
units where $c = \hbar = 1$ throughout the document unless noted otherwise} 
provided good agreement with $\chi$PT predictions. 

These LETs \cite{kr,namlu,namsh,vz,schkoch} are not
applicable for $Q^{2} \gg \Lambda_{QCD}^3/m_\pi$, where $\Lambda_{QCD} \sim 
200-300$
MeV is the QCD scale parameter. In the case of asymptotically large momentum
transfers ($Q^2 \to \infty$) perturbative QCD (pQCD) factorization 
techniques \cite{pobpolstrik,efremrad,lepagebrodsky} have been used to obtain
predictions for cross section amplitudes and axial form factors near threshold.
In these factorization techniques, 
`hard' ($Q^2 \gg \Lambda^2_{QCD}$) and `soft' ($k \sim \Lambda_{QCD}$) momentum 
contributions to the scattering amplitude can be separated cleanly and each 
contribution can be theoretically calculated using pQCD and LETs, respectively.
Here, $k$ is the momentum of the virtual photon. 

Recently, Braun \textit{et al}.~\cite{braunlcsr,braunrev} suggested 
a method to extract the generalized form factors, $G_1^{\pi N}(Q^2)$ and 
$G_2^{\pi N}(Q^2)$, 
for $1 < Q^2 < 10$ GeV$^2$ using light cone sum rules (LCSR).
The transition matrix elements of the electromagnetic interaction, $J_\mu$, 
can be
written in terms of these form factors at threshold:
\begin{widetext}
  \begin{eqnarray}
    \langle N(P') \pi(k) | J_\mu | p(P) \rangle  = 
     - \frac{i}{f_\pi} \bar{N} \gamma_5 
    \left[ (\gamma_\mu q^2 - q_\mu \slashed{q})\frac{G_1^{\pi N}(Q^2)}{m_N^2} 
     - \frac{i\sigma_{\mu\nu} q^\nu}{2m_N} G_2^{\pi N}(Q^2)\right] p. 
  \end{eqnarray}  
\end{widetext}
Here, $N(P')$ and $p(P)$ are spinors for the final and initial nucleons 
with momenta $P'$ and $P$, respectively, $m_N$ is the mass of the nucleon,
$f_\pi$ is the pion decay constant and $q$ is the 4-momentum of the virtual photon. 
Since the pion is a negative parity particle
and the electromagnetic current is parity conserving, the $\gamma_5$
matrix is present to conserve the overall parity of the reaction.

These form factors are directly related to the pion-nucleon $s$-wave
multipoles $E_{0+}$ and $L_{0+}$ \cite{braunlcsr,braunrev}
\begin{eqnarray}\label{eqn:ffmultrelation1}
  E_{0+} & = & \frac{\sqrt{4\pi\alpha}}{8\pi f_\pi} 
  \sqrt{\frac{(2m_N + m_\pi)^2 + Q^2}{m_N^3(m_N + m_\pi)^3}} \nonumber\\
  & & \times
  \left( Q^2 G_1^{\pi N} - \frac{m_N m_\pi}{2}G_2^{\pi N}\right)\\
  L_{0+} & = & \frac{\sqrt{4\pi\alpha}}{8\pi f_\pi} 
  \frac{m_N |\omega^{th}_\gamma|}{2}
  \sqrt{\frac{(2m_N + m_\pi)^2 + Q^2}{m_N^3(m_N + m_\pi)^3}} \nonumber \\
  & & \times
  \left( G_2^{\pi N} + \frac{2 m_\pi}{m_N}G_1^{\pi N}\right).
\end{eqnarray} 
Here, $\alpha$ is the electromagnetic coupling constant and $\omega_{\gamma}^{th}$ is the 
virtual photon energy at
threshold in the c.m.~frame and is given by the following relation:
\begin{equation}
  \omega^{th}_\gamma = \frac{m_\pi(2m_N + m_\pi) - Q^2}{2(m_N + m_\pi)}.
\end{equation}
In general, $E_{l\pm}$, $M_{l\pm}$, and $L_{l\pm}$ describe the electric, magnetic
and longitudinal multipoles, respectively. Here, $l$ describes the total orbital angular
momentum of the pion relative to the nucleon and $\pm$ is short for $\pm \frac{1}{2}$
so that the total angular momentum of the $\pi N$ system is $l \pm \frac{1}{2}$. 

Additionally, the sum rules can be extended to
the $Q^2 \sim 1$ GeV$^2$ regime and the LETs are recovered 
to $O(m_\pi)$ accuracy by including contributions from
semi disconnected pion-nucleon diagrams \cite{braunrev}. 
This approach provides a connection between the low and
high $Q^2$ regimes. Predictions for the axial form factor and the generalized
form factors are also obtained in this approach.  

In the low $Q^2 < 1$ GeV$^2$ regime and the chiral limit $m_\pi \to 0$, the 
LET $s$-wave multipoles at threshold can be written as \cite{schkoch}:
\begin{eqnarray}\label{eqn:letmult}
  E_{0+} & = &\frac{\sqrt{4\pi\alpha}}{8\pi}\frac{Q^2\sqrt{Q^2+4m_N^2}}{m_N^3f_\pi}
    G_1^{\pi N} ,\\
  L_{0+} & = &\frac{\sqrt{4\pi\alpha}}{32\pi}\frac{Q^2\sqrt{Q^2+4m_N^2}}{m_N^3f_\pi}
    G_2^{\pi N}.
\end{eqnarray}
$G_1^{\pi N}$ and $G_2^{\pi N}$ can be written in terms of the electromagnetic
form factors for the neutral pion-proton $\pi^0 p$ channel in this approximation:
\begin{eqnarray}\label{eqn:letff}
  \frac{Q^2}{m_N^2} G_1^{\pi^0 p} & = &\frac{g_A}{2} \frac{Q^2}{(Q^2 + 2m_N^2)} G_M^p, \\
  G_2^{\pi^0 p} & = & \frac{2g_A m_N^2}{Q^2 + 2m_N^2} G_E^p.
\end{eqnarray}
In the above equations, $G_M^p$ and $G_E^p$ are the Sachs electromagnetic form
factors of the proton and $g_A$ is the axial coupling constant obtained from weak 
interactions. Also, for the charged pion-neutron $\pi^+ n$ channel, the generalized
form factors can be written as:
\begin{eqnarray}\label{eqn:letffneutron}
  \frac{Q^2}{m^2_N} G_1^{\pi^+ n} & = & \frac{g_A}{\sqrt{2}} \frac{Q^2}{(Q^2 + 2m_N^2)} G_M^n + \frac{1}{\sqrt{2}} G_A, \\
  G_2^{\pi^+ n} & = &\frac{2\sqrt{2}g_A m_N^2}{Q^2 + 2m_N^2} G_E^n.
\end{eqnarray}
Here, $G_M^n$ and $G_E^n$ are the electromagnetic form factors of the neutron. 
Additionally, $G_A$ is the axial form factor that is induced by the charged 
current and its contribution comes from the Kroll-Ruderman term \cite{kr}. 

These generalized form factors, $G_1^{\pi N}$ and $G_2^{\pi N}$, 
can be described as overlap integrals 
of the nucleon and the pion-nucleon wave functions. The wave
function of the pion-nucleon system at threshold is related to the nucleon
wave function without the pion by a chiral rotation in the spin-isospin space 
\cite{pobpolstrik,braunlcsr}. The measurement of these form factors for
pion electroproduction is in essence the measurement of the 
overlap integrals of the rotated and non-rotated nucleon wave functions,
which are not accessible in elastic form factor measurements. This
information complements our understanding of the various components
of the nucleon wave function (quarks and gluons) and the theory of strong interactions. 
Additionally, it provides insight into chiral symmetry and 
its violation in reactions at increasing $Q^2$. 

The generalized form factor for the charged pion-neutron $G_1^{\pi^+ n}(Q^2)$ and
the axial form factor $G_A(Q^2)$ had been measured near threshold for 
$Q^2 \sim 2 - 4.2$ GeV$^2$ \cite{kijun2012}. 
In this paper, we describe the measurement of the differential cross sections
and the extraction of the $s$-wave amplitudes for 
the neutral pion electroproduction process, $e p \to e p \pi^0$, for
$Q^2 \sim 2 - 4.5$ GeV$^2$ near threshold, \emph{i.e.}, $W \sim 1.08 - 1.16$ GeV.
From these cross sections, the generalized
form factors $G_1^{\pi^0 p}(Q^2)$ and $G_2^{\pi^0 p}(Q^2)$ were extracted and
compared with the theoretical calculations of Refs.~\cite{braunrev} and 
\cite{schkoch}. 


\section{Kinematic Definitions and Notations\label{sec:kindef}}

\begin{figure}
  \centering
  \includegraphics[width=8cm,height=5cm]{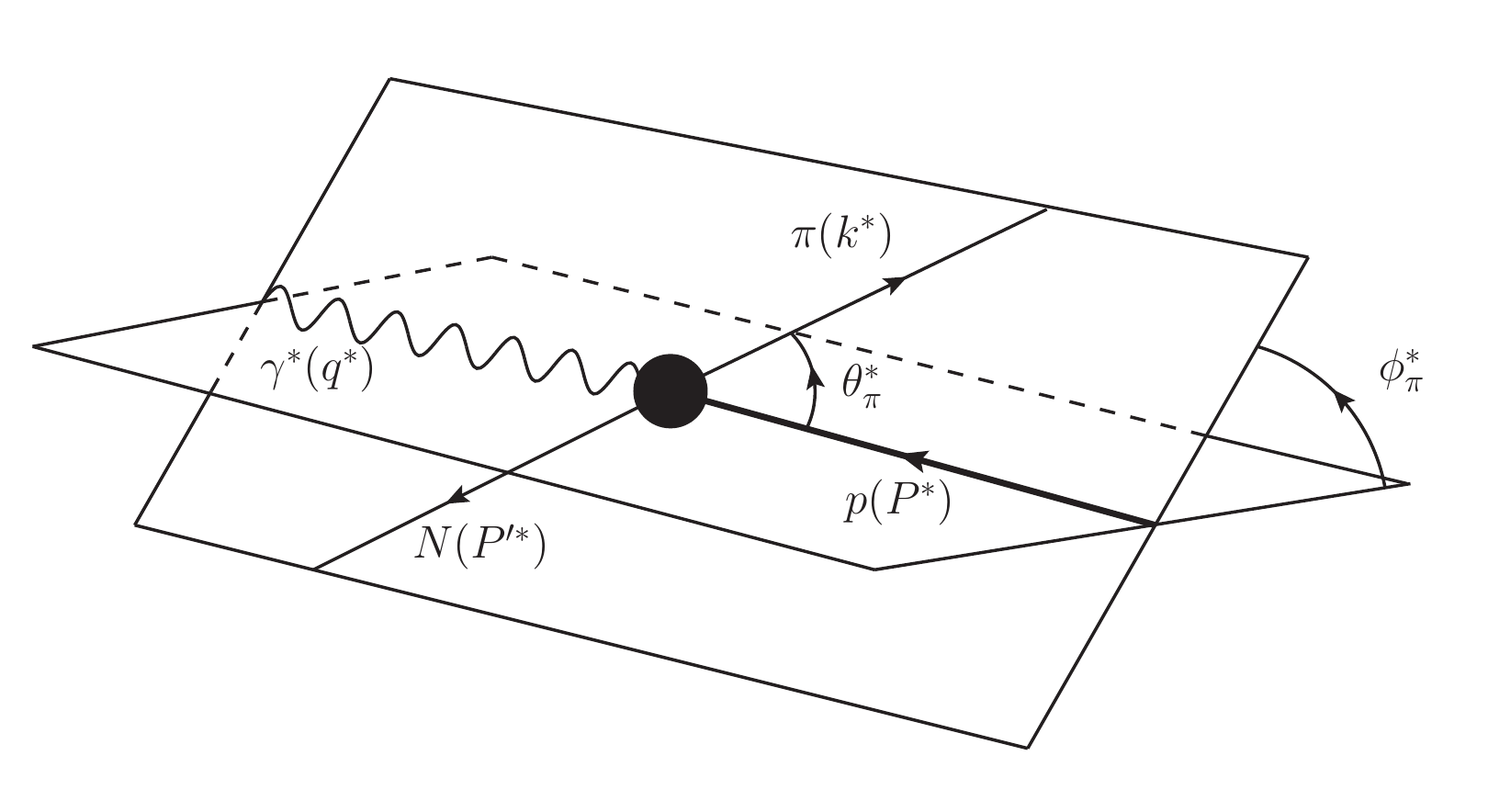}
  \caption{\label{fig:fig1} Neutral pion electroproduction in the center 
  of mass frame.}
\end{figure}

The neutral pion reaction
\begin{equation}
  e(l) + p(P) \to e(l') + p(P') + \pi^0 (k)
\end{equation}
is shown schematically in the virtual photon-proton center of mass frame in
Fig.~\ref{fig:fig1}. Here, $l=(E_e, \mathbf{p_e})$, 
$l'=(E'_{e}, \mathbf{p'_{e}})$, $P=(m_p, \mathbf{0})$ and 
$P'=(E'_{p}, \mathbf{p'_{p}})$ are the initial and final electron and proton
4-momenta in the lab frame and $k=(E_{\pi}, \mathbf{p_{\pi}})$ is the 
4-momentum of the emitted pion. Also, $m_p$ refers to the mass of the proton. 
It is assumed that the incident electron interacts 
with the target proton via exchange of a single virtual photon with 4-momentum
$q = l - l' = (\omega, \mathbf{q})$. In this 
approximation, it is also assumed that the electron mass is negligible 
($m_e \approx 0$).
The two important kinematic invariants of interest are
\begin{eqnarray}
  Q^2 \equiv -q^2 = -\omega^2 + |\mathbf{q}|^2 = 4E_e E'_e \sin^2 (\theta'_e/2)\nonumber\\
  s = W^2 = (q+P)^2 = m_{p}^2 + 2\omega m_{p} - Q^2.
\end{eqnarray}
Here, 
$\theta'_e$ is the polar angle of the scattered electron in the lab frame. 

The five-fold differential cross section for the reaction can be written
in terms of the cross section for the subprocess 
$\gamma^* p \to p\pi^0$ \cite{amaldi}, which depends only on the matrix elements 
of the hadronic interaction:
\begin{equation}\label{eqn:diffcross1}
\frac{d^5\sigma}{dE^{\prime}_{e} d\Omega^{\prime}_{e} d\Omega^{*}_{\pi}} = 
\Gamma \frac{d^2\sigma_{\gamma^{*}p}}{d\Omega^{*}_{\pi}}.
\end{equation}
Here, $d\Omega^{\prime}_{e} = d\cos\theta^{\prime}_{e} d\phi^{\prime}_{e}$ is
the differential solid angle for the scattered electron in the lab frame and
$d\Omega^{*}_{\pi} = d\cos\theta^{*}_{\pi} d\phi^{*}_{\pi}$ is the differential
solid angle for the pion in the virtual photon-proton ($\gamma^* p$) center of 
mass frame. The azimuthal angle $\phi^*_\pi$ is determined with respect to
the plane defined by the incident and scattered lepton \cite{drechtiat}.
The factor $\Gamma$ represents the \textit{virtual photon flux}. In 
the Hand convention \cite{amaldi} it is 
\begin{equation}\label{eqn:vgflux}
\Gamma = \frac{\alpha}{2\pi^{2}}\frac{E^{\prime}_e}{E_e} 
\frac{W^2 - m^2_p}{2m_pQ^2}\frac{1}{1-\varepsilon},
\end{equation}
which depends entirely on the
matrix elements of the leptonic interaction and contains the 
\textit{transverse polarization} of the virtual photon
\begin{equation}
  \varepsilon = \left ( 1 + 2\frac{|\mathbf{q}|^2}{Q^2}\tan^{2}
  \frac{\theta_{e}^{\prime}}{2}\right)^{-1}.
\end{equation} 

For unpolarized beam and target the reduced cross section from 
Eq.~(\ref{eqn:diffcross1}) can be expanded in terms of the hadronic structure 
functions:
\begin{eqnarray}\label{eqn:diffeqsf}
\frac{d\sigma_{\gamma^{*}p}}{d\Omega^*_\pi} & = & 
\frac{|\mathbf{p_\pi^*}|}{K}
\left[\frac{d\sigma_T}{d\Omega^*_\pi} + \varepsilon \frac{d\sigma_L}{d\Omega^*_\pi} 
+\varepsilon\frac{d\sigma_{TT}}{d\Omega^*_\pi} \cos2\phi^{*}_{\pi} \right.\nonumber\\
& & + \left.\sqrt{2\varepsilon(\varepsilon+1)} \frac{d\sigma_{LT}}{d\Omega^*_\pi}
\cos\phi^{*}_{\pi} \right].
\end{eqnarray}
Here, 
$\mathbf{p_\pi^*}$ is the pion momentum and $K = (W^2 - m_p^2)/2W$ is the
photon equivalent energy in the c.m.~frame of the subprocess $\gamma^*p \rightarrow p \pi^0$. 
Additionally, $\sigma_T + \varepsilon\sigma_L$, $\sigma_{LT}$ and $\sigma_{TT}$ 
are the structure
functions that describe the transverse, longitudinal, longitudinal-transverse
interference, and transverse-transverse interference components of the 
differential cross section. 

Each of these structure functions contain the $\cos\theta^*_\pi$ dependence and
can be parameterized in terms of the multipole amplitudes 
$E_{l\pm}$, $M_{l\pm}$ and $S_{l\pm}$ that describe the electric, magnetic
and scalar multipoles, respectively. 
The scalar multipoles $S_{l\pm}$ can be written in terms of the longitudinal 
multipoles $L_{l\pm} = \frac{\omega^*}{|\mathbf{q^*}|}S_{l\pm}$, where $\omega^*$ and
$\mathbf{q^*}$ are the energy and 3-momentum of the virtual photon in the c.m.~frame,
respectively \cite{drechtiat}.


\section{Experiment}

\begin{figure}
  \subfigure[]{
    \includegraphics[width=6cm]{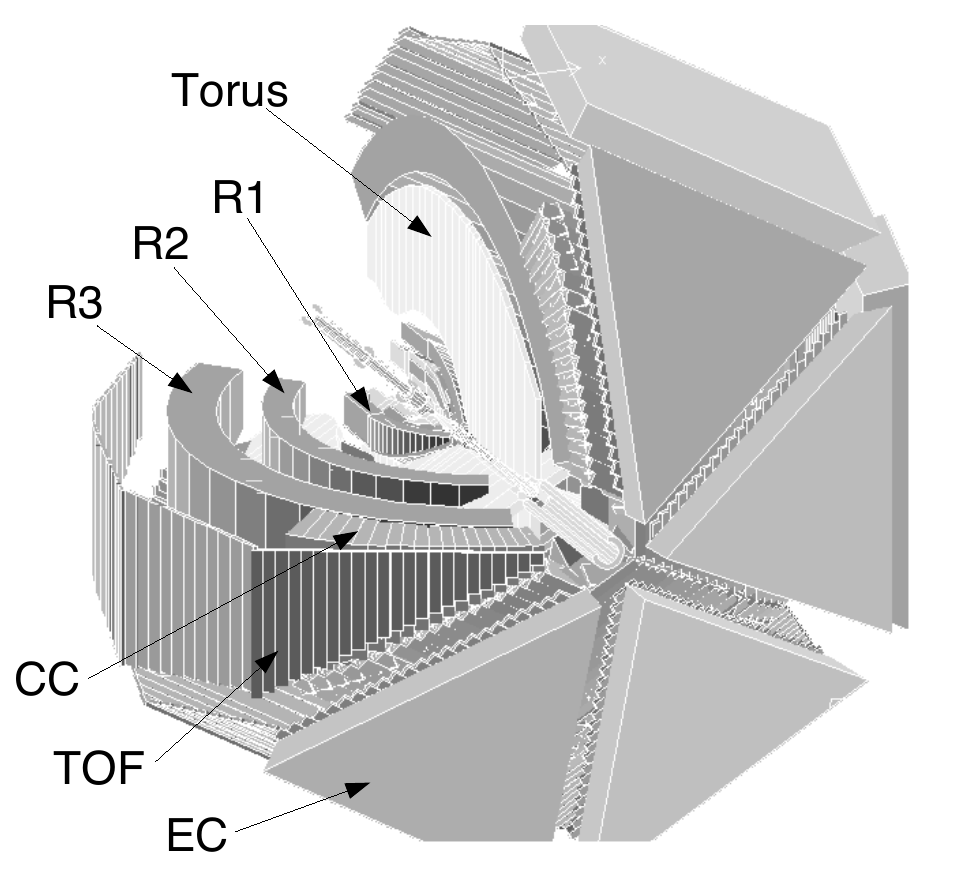}  
    \label{fig:fig2a}
  }
  \subfigure[]{
    \includegraphics[width=5cm]{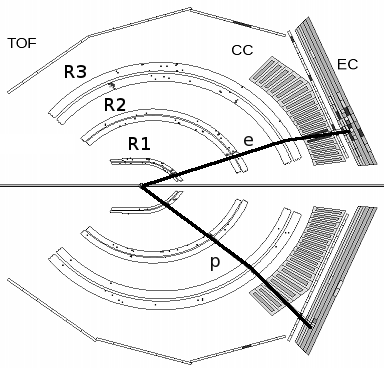}
    \label{fig:fig2b}
  }
  \caption{\subref{fig:fig2a} A three-dimensional view of CLAS showing 
  the superconducting coils of the torus, the three regions of drift chambers 
  (R1-R3), the \v{C}erenkov counters, the time-of-flight system, and
  the electromagnetic calorimeters. The positive
  \textbf{\^z}-axis is out of the page along the symmetry axis. 
  \subref{fig:fig2b} A schematic view of a typical near threshold event
  showing the reconstructed electron and proton tracks with the corresponding
  detector hits in two opposite CLAS sectors. The $\pi^0$ is reconstructed
  using the missing mass technique as discussed in the text.}
\end{figure}

The near threshold reaction $ep \to ep\pi^0$ was studied using the CEBAF Large 
Acceptance Spectrometer (CLAS) in Jefferson Lab's 
Hall-B \cite{clas}. Fig.~\ref{fig:fig2a}
shows the detector components that comprise CLAS. Six superconducting coils of
the torus divide CLAS into six identical sectors and produce a toroidal 
magnetic field in the azimuthal direction around the beam axis. Each of the six 
sectors contain three regions of drift chambers (R1, R2, and R3) to track 
charged particles and to reconstruct their momentum \cite{clas_dc}, 
scintillator counters for identifying particles based on 
time-of-flight (TOF) information \cite{clas_tof}, \v{C}erenkov counters (CC) to 
identify electrons \cite{clas_cc}, and electromagnetic counters (EC) to 
identify electrons and neutral particles \cite{clas_ec}. The CC and EC are
used for triggering on electrons and provide a mechanism to separate charged
pions and electrons.
With these six sectors, CLAS provides a large solid angle coverage
with typical momentum resolutions of about $0.5\% - 1.0\%$  
depending on the kinematics \cite{clas}. 

A 5.754 GeV electron beam with an average intensity of
7 nA was incident on a 5 cm long liquid hydrogen target, which was placed 4 cm 
upstream of the CLAS center. Fig.~\ref{fig:fig2a} shows 
the electron beam entering CLAS from the top left and exiting from the bottom right 
through the symmetry axis. A small non-superconducting magnet (minitorus) surrounded 
the target and generated a toroidal field to shield the R1 drift chambers from low 
energy electrons of high intensity. These electrons originated primarily from the 
M{\o}ller scattering process. The data used in this experiment were collected from 
October 2001 to January 2002 and the integrated luminosity was about 
$0.28$ fb$^{-1}$.
The electron beam energy of 5.754 GeV as determined in this experiment agrees 
within 6 MeV with an independent measurement in Hall A \cite{hallanim}.


\section{Analysis}

At the start of this analysis, a cut of $W < 1.3$ GeV is applied to focus our
events only in the kinematic region of interest.
In this analysis the scattered electrons and protons are detected using CLAS
and the $\pi^0$ is reconstructed using 4-momentum conservation.
A typical event for this experiment is shown in Fig.~\ref{fig:fig2b}. 


\subsection{Particle Identification: Electron}\label{sec:epid}

\begin{figure}
  \subfigure[]{
    \includegraphics[width=8cm]{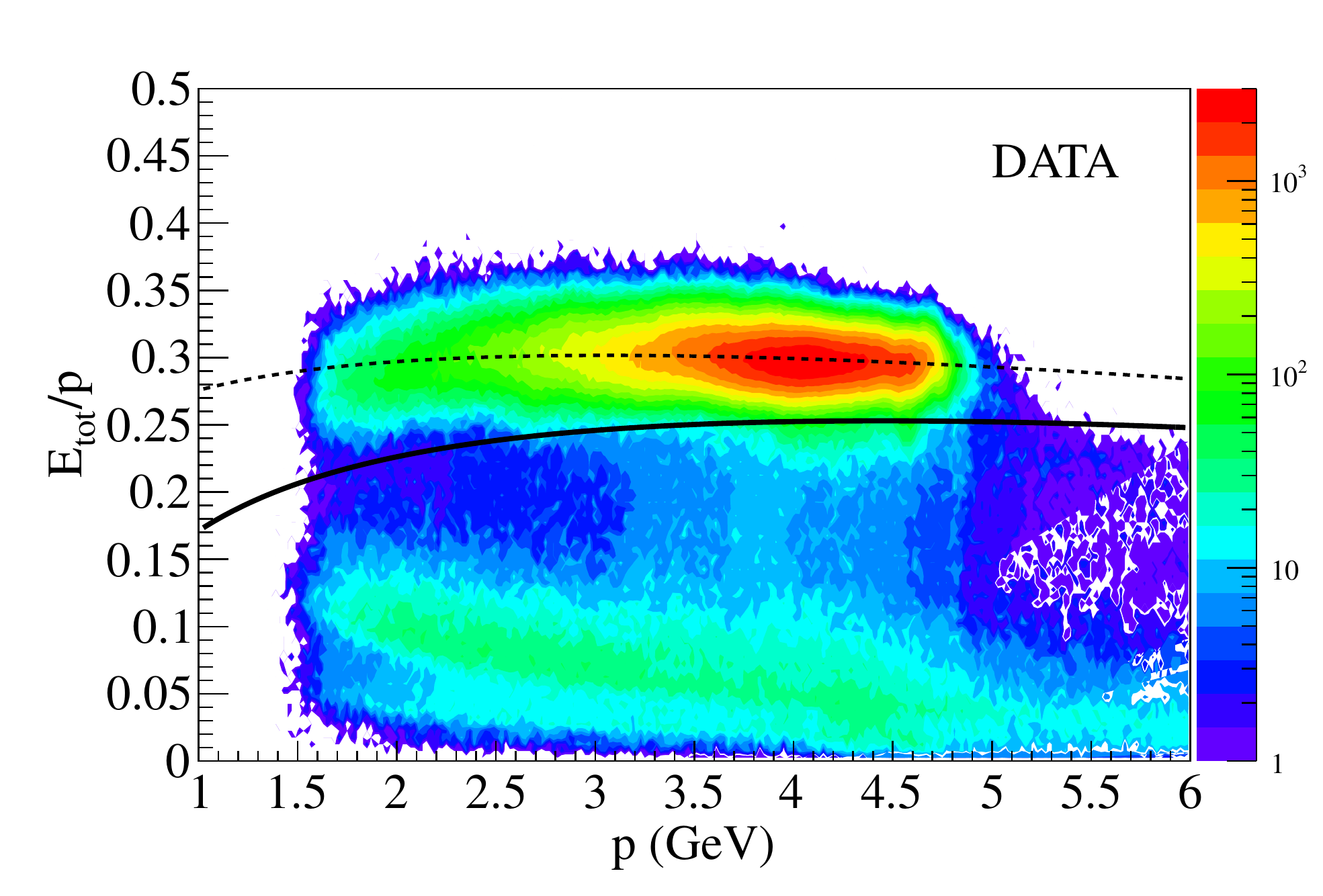}
    \label{fig:fig3a}
  }
  \subfigure[]{
    \includegraphics[width=8cm]{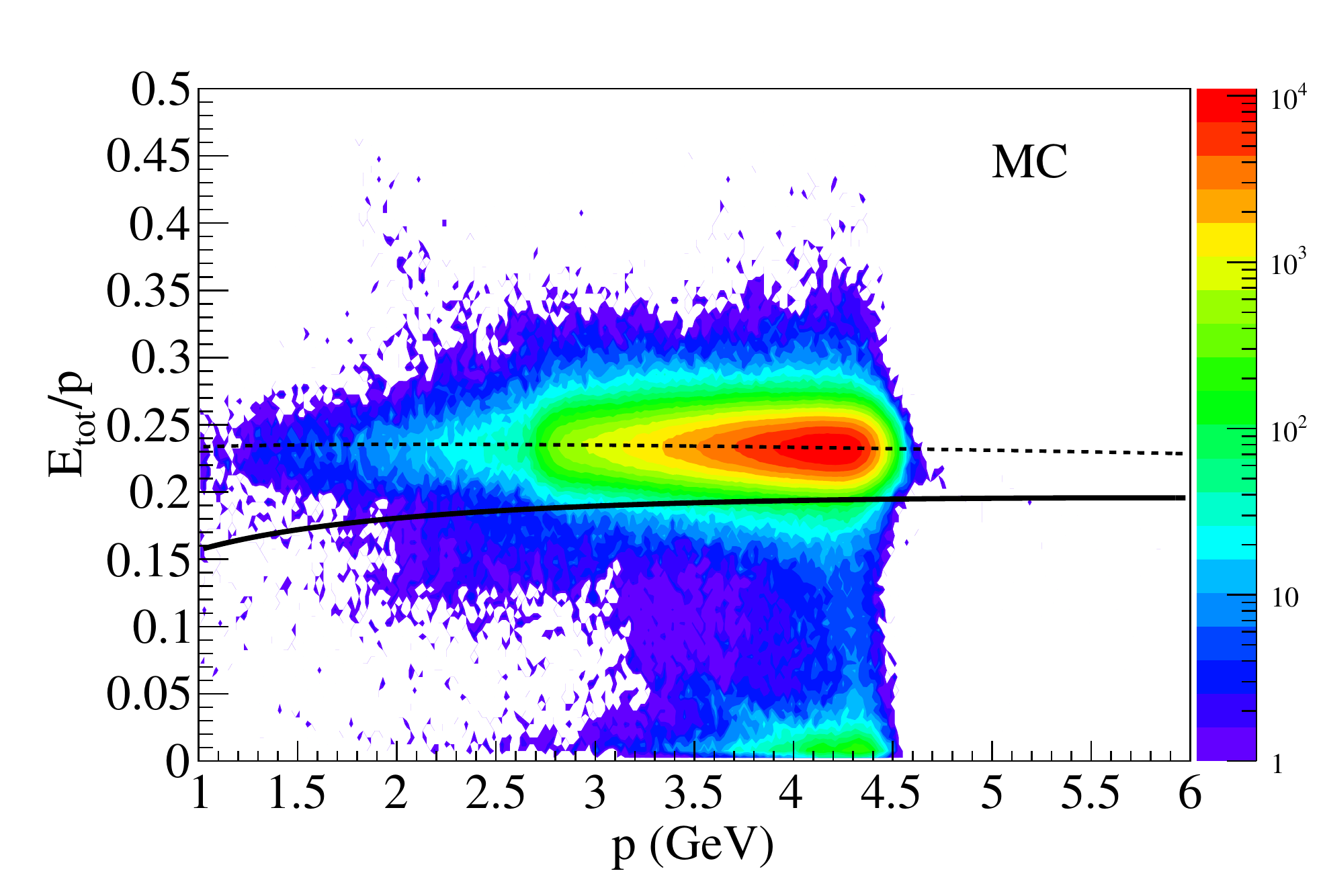}
    \label{fig:fig3b}
  }
  \caption{(Color online) EC sampling fraction as a function of electron 
  momentum for one of the CLAS sectors for \subref{fig:fig3a} Data 
  and \subref{fig:fig3b} Monte Carlo (MC) simulation. The dashed lines show the 
  parameterized mean and the solid line indicates the $3\sigma$ cut.}
  \label{fig:ec}
\end{figure}

The scattered electrons in the final state of the reaction are detected by
requiring geometrical coincidence between the \v{C}erenkov counters and the 
electromagnetic calorimeter in the same sector. The momentum
of the electrons is reconstructed using the drift chambers. Using the 
energy deposited in the EC and the momentum, the electrons are isolated
from most of the minimum ionizing particles (MIPs), \textit{e.g.}, pions, contaminating 
the electron spectra.

As electrons pass through the EC, they shower with a total energy deposition 
$E_{tot}$ that is proportional to their momenta $p$. The sampling fraction
energy $E_{tot}/p$ is plotted as a function of momentum for each sector after
applying all the other electron identification cuts. Fig.~\ref{fig:ec} shows
this distribution for one of the CLAS sectors for experimental and Monte Carlo
simulated events. In the figure, one can note the
MIPs contamination near the smaller values of $E_{tot}/p$. This contamination
is significantly larger in data than in simulated events. 
The electrons are concentrated near $E_{tot}/p \approx 0.3$.
Ideally they should not show any dependence on momentum, albeit a slight momentum
dependence is visible in the data. This dependence is parameterized and a cut of 
$3\sigma$ is applied as shown in the figure. The MIP events are well
separated from the electrons below the $3\sigma$ cut.


\subsection{Particle Identification: Proton}

\begin{figure}
\centering
  \subfigure[]{
    \includegraphics[width=8cm]{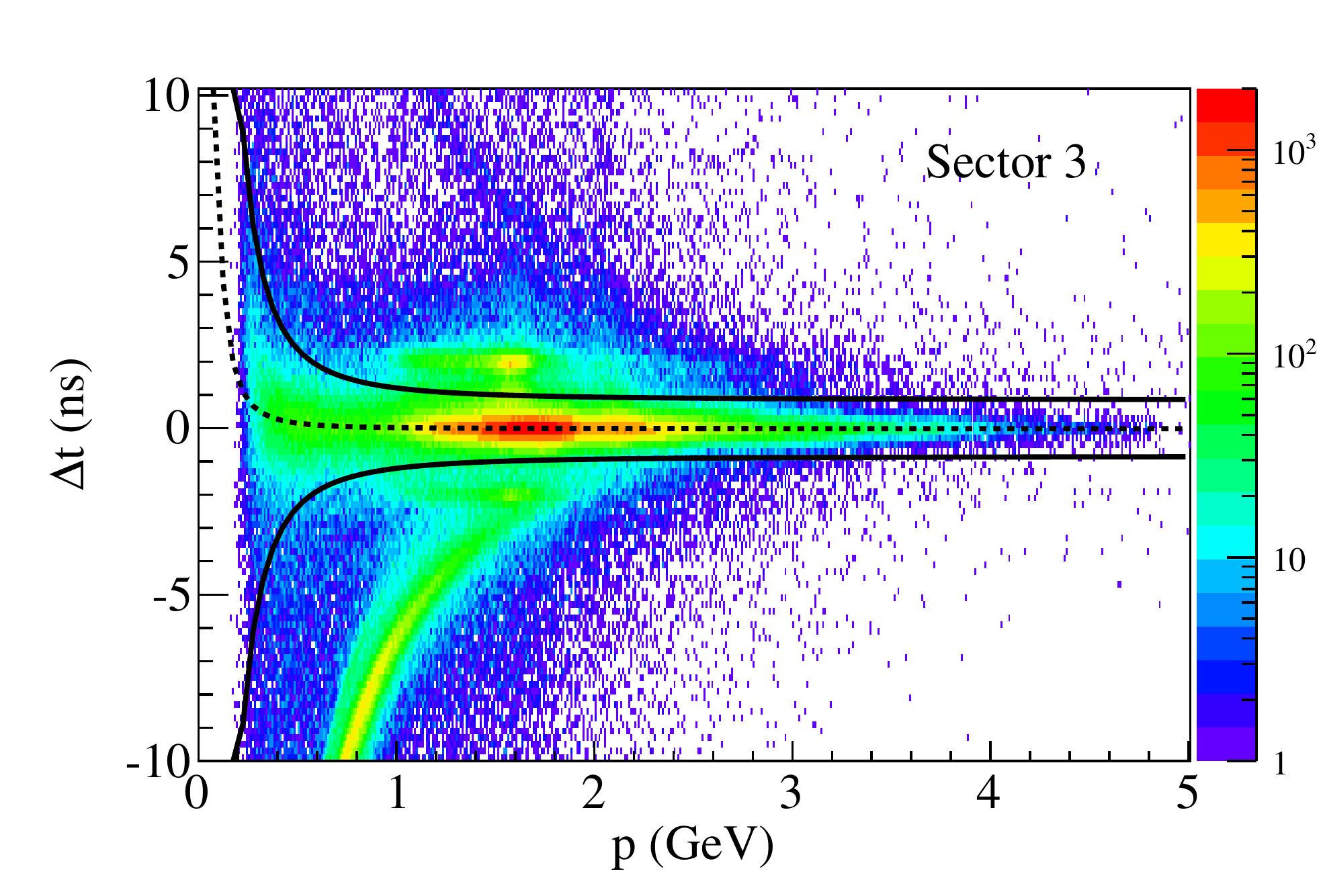}
    \label{fig:fig4a}}
  \subfigure[]{
      \includegraphics[width=8cm]{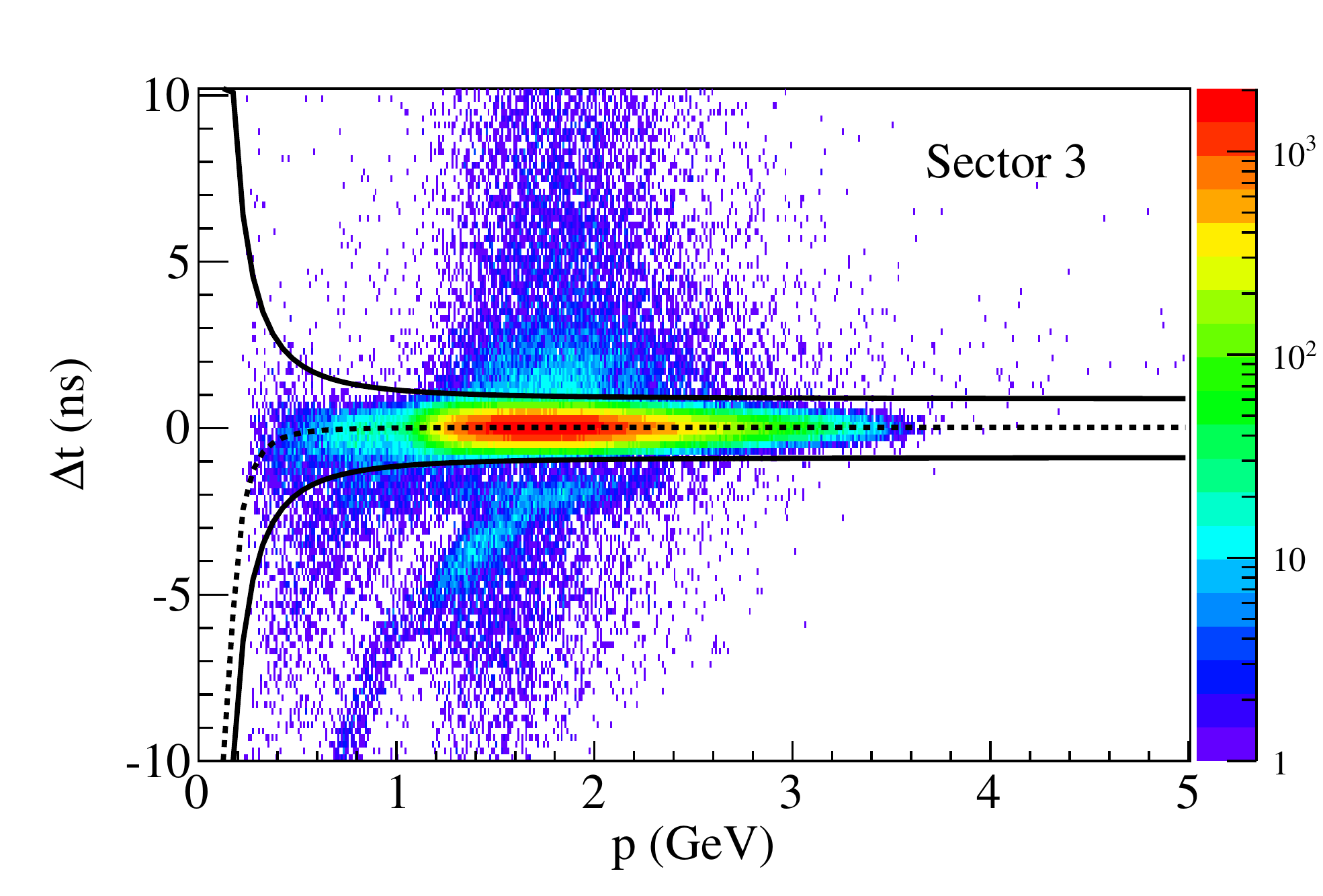}
      \label{fig:fig4b}}
  \caption{(Color online) $\Delta t$ as a function of $p$. The curves show
    the $\pm 3.5\sigma$ cut (solid lines) from the mean fit 
    (dashed line) for one of the CLAS sectors for \subref{fig:fig4a}
    experimental and \subref{fig:fig4b} Monte Carlo simulated events.
  }
\end{figure}

The recoiled protons are identified using the measured momentum and the timing 
information obtained from the TOF counters. A track is selected as a proton 
whose measured time is closest to that expected of a real proton, \textit{i.e.},
\begin{equation}\label{eqn:dt}
  \Delta t = t_{meas} - t_{calc} = \left( t_{TOF} -  t_{tr}\right ) - 
  \frac{l}{\beta_{calc} c} .
\end{equation}
In the above equation, $t_{TOF}$ is the time measured from the 
TOF counters, $l$ is the distance from the target center to the TOF paddle,
and $t_{tr}$ is the event start time calculated from the 
electron hit time from the TOF traced back to the target position. 
Also, in Eq.~(\ref{eqn:dt})
$\beta_{calc} = p/\sqrt{M^2_{pdg} + p^2}$, where $\beta_{calc}$ is computed using
the PDG \cite{pdg} value of the mass of the proton $M_{pdg}$ and the 
momentum of the track $p$.

Figs.~\ref{fig:fig4a} and \subref{fig:fig4b} show the experimental and
simulated event distributions, respectively, of $\Delta t$ as a function of $p$ 
for one of the CLAS sectors. The protons are centered around $\Delta t = 0$ ns 
and have a slight momentum dependence for $p < 1$ GeV.
The dashed lines indicate the parameterized mean
of the distributions and the solid lines indicate the $\pm3.5\sigma$ cut applied 
to select the protons.


\subsection{Fiducial Cuts and Kinematic Corrections}

For perfect beam alignment, the incident electron beam is expected to be centered at 
$(X_{beam}, Y_{beam}) = (0,0)$ cm at the target. But due to misalignments, the electron
beam was actually at $(X_{beam}, Y_{beam}) = (0.090, -0.345)$ cm.
This misalignment of the beam-axis is corrected for each sector, which also subsequently
changes the reconstructed $z$-vertex positions of the electron and proton tracks. 
The details of this correction are described in previous works \cite{ungaroprl,kijunprc2008}. 
A cut of $z \in (-8.0,-0.8)$ cm is placed on the $z$-vertex to isolate events from within the 
target cell. 

The measured angles and momenta of the electrons and protons are corrected using the 
same method as used in previous analyses \cite{ungaroprl,kijunprc2008}.

The electrons start to lose energy as they enter the electromagnetic calorimeter. 
When the electrons shower near the edge of the calorimeter, their shower is not fully 
contained and so their energies cannot be properly reconstructed. 
As such, a fiducial cut is applied to remove these events.  

Electrons give off \v{C}erenkov light in the CC, which is collected in the PMTs on either 
side of the counters in each sector. 
Inefficient regions in the CC are isolated by removing those regions where the average number of 
photo-electrons $\langle Nphe \rangle < 5$. This cut results in keeping all events
that lie in regions where the CC efficiency is about 99\% \cite{clas_cc}.

\begin{figure}
  \centering
  \includegraphics[width=8.5cm]{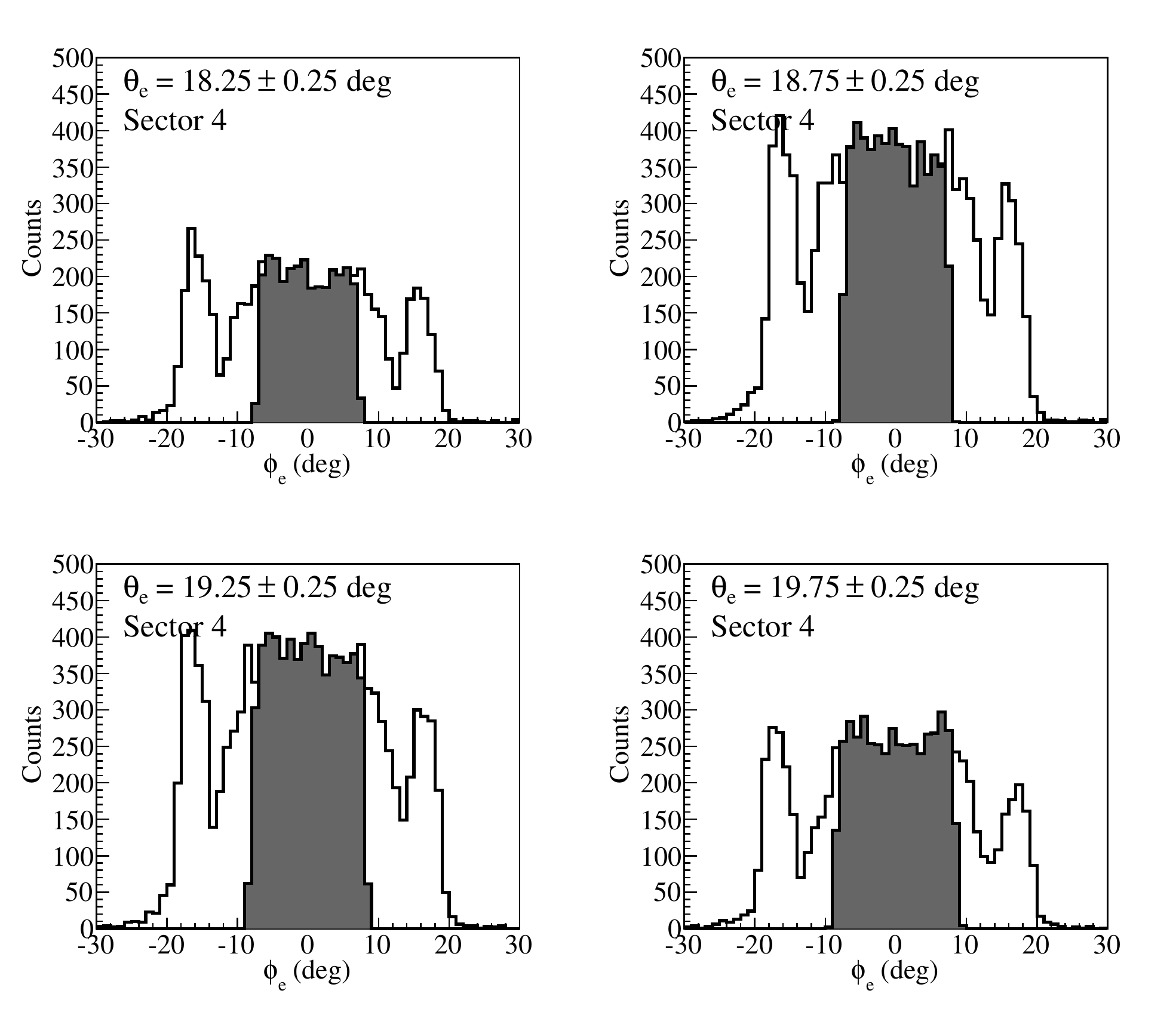}
  \caption{\label{fig:fig5} Electron $\phi_e$  
	distribution for CLAS Sector 4 for $p_e = 4.1 \pm 0.1$ GeV shown for different $\theta_e$
	slices. The unshaded curves show $\phi_e$ distribution after electron selection
	and the shaded curves show the $\phi_e$ distribution after applying electron DC 
	fiducial cuts.}
\end{figure}

To deal with edges and holes in the drift chambers, and to remove dead or inefficient wires, a
fiducial cut for both electrons and protons is applied. Regions of non-uniform 
acceptance in the azimuthal angle $\phi$ resulting from these attributes are isolated on 
a sector-by-sector basis as a function of the electron's momentum $p_e$ and polar angle 
$\theta_e$. For the electron, at fixed $p_e$ and $\theta_e$, one expects the angular 
distribution
to be symmetric in $\phi_e$ and relatively flat. Empirical cuts are applied to select these 
regions of relatively flat $\phi_e$ as shown in Fig.~\ref{fig:fig5} for electrons with
$p = 4.1 \pm 0.1$ GeV for different slices of $\theta_e$ and one of the CLAS sectors. 
The same cuts are applied to both experimental and simulated events. 

\begin{figure}
	\centering
	  \includegraphics[width=8cm]{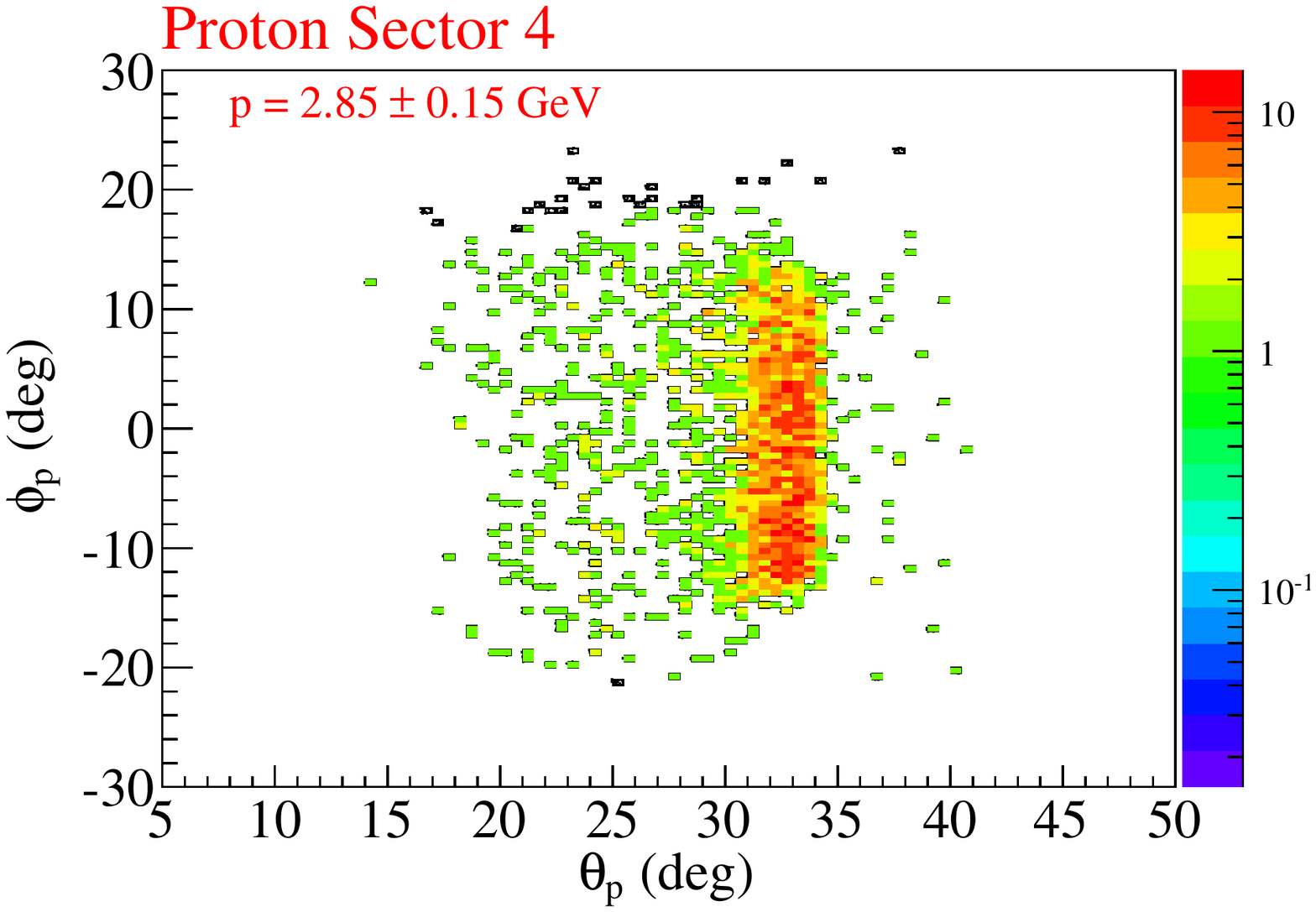}
	\caption{(Color online) \label{fig:fig6} Proton $\phi_p$ vs.~$\theta_p$ distribution for 
	CLAS Sector 4 for $p_p = 2.85 \pm 0.15$ GeV. Rejected tracks are shown in black.} 
\end{figure}

As for electrons, a fiducial cut on the proton's azimuthal angle $\phi_p$ as a function of
its momentum $p_p$ and polar angle $\theta_p$ is applied. However, the edges of the $\phi_p$ 
distributions are asymmetric for different slices of $\theta_p$. 
The upper and lower bounds on $\phi_p$ are extracted and parameterized as a function of 
$\theta_p$ and $p_p$. The result of this cut for one of the CLAS sectors is shown in 
Fig.~\ref{fig:fig6}.

\subsection{Background Subtraction and $\pi^0$ Identification}\label{sec:bhsub}

The neutral pion in the final state is reconstructed using energy and 
momentum conservation constraint. To do so, we use the conservation of 4-momentum
and look at the missing mass squared distribution of the detected particles (\textit{i.e.},
the electron and the proton):
\begin{equation}
	M_X^2 (ep) = (l + P - l' - P')^2.
\end{equation}
Here, $l$, $P$, $l'$ and $P'$ are 4-momenta of the incident and scattered
particles as described in Section \ref{sec:kindef}.

\begin{figure}
	\centering
	\subfigure[]{	
	  \includegraphics[width=4cm]{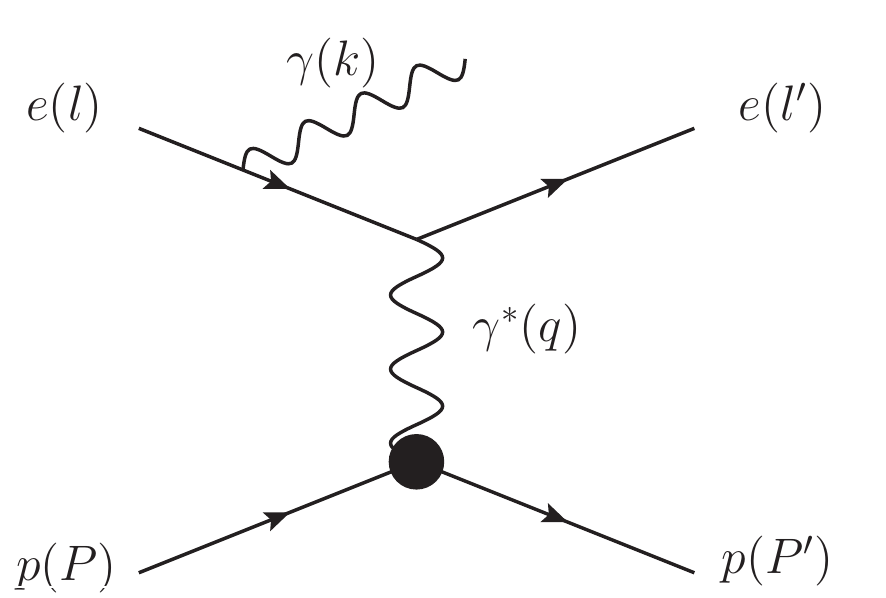}
	  \label{fig:fig7a}
	}
	\subfigure[]{	
	  \includegraphics[width=4cm]{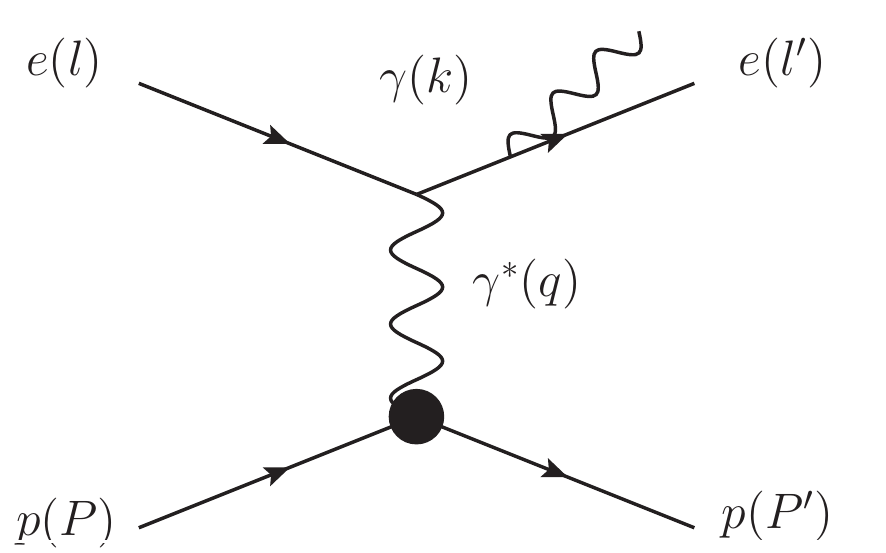}
	  \label{fig:fig7b}
	}
	\caption{The Bethe-Heitler process $ep \to ep\gamma$ diagrams for	
	\subref{fig:fig7a} a photon emitted from an incident electron
	(pre-radiation)
	and for \subref{fig:fig7b} a photon emitted from a scattered electron
	(post-radiation). \label{fig:bhprepostrad}}	
\end{figure}

There are several difficulties in the analysis in the 
near threshold region. In this region, the pion electroproduction cross section 
goes to zero; so, the statistics are very low. Also, a major source of 
contamination to the neutral pion signal near threshold is the elastic 
Bethe-Heitler process $ep \to ep\gamma$. The two dominating Feynman diagrams for this
process are shown in Fig.~\ref{fig:bhprepostrad}. Fig.~\ref{fig:fig7a} shows the 
diagram with a pre-radiated photon (emission from an incident electron) and
Fig.~\ref{fig:fig7b} shows the diagram with a post-radiated photon 
(emission from a scattered electron). These photons are emitted approximately in the
direction of the incident and scattered electron, respectively \cite{schiffpa,entpa}.
When these photons are emitted, the
incident and scattered electrons lose energy. This feature of the 
Bethe-Heitler process can be exploited to our benefit.

For the elastic process $ep\to ep$, the proton angle can be computed 
independently of the incident or scattered electron energies:
\begin{eqnarray}
	\tan \theta^p_1 & = & \frac{1}{\left(1 + \frac{E'}{m_p - E' \cos\theta_e'} 
	 \right) \tan\frac{\theta_e'}{2}} \\
	\tan \theta^p_2 & = & \frac{1}{\left(1 + \frac{E}{m_p}\right) 
	 \tan\frac{\theta_e'}{2}}.
\end{eqnarray}
Here, $\theta^p_1$ and $\theta^p_2$ are the proton angles computed independently
of the incident or scattered electron energies, respectively. Also, 
$\theta_e'$ is the angle of the scattered electron in
the lab frame, and $E$ and $E'$ are the energies of the incident and 
scattered electron, respectively. We can calculate these angles for each
event and look at its deviation ($\Delta\theta_{1,2}^p$) from the 
measured value ($\theta^p_{meas}$):
\begin{equation}
	\Delta\theta_{1,2}^p \equiv \theta_{1,2}^p - \theta^p_{meas}.
\end{equation} 

\begin{figure}
	\centering
	\subfigure[]{
	  \includegraphics[width=8cm]{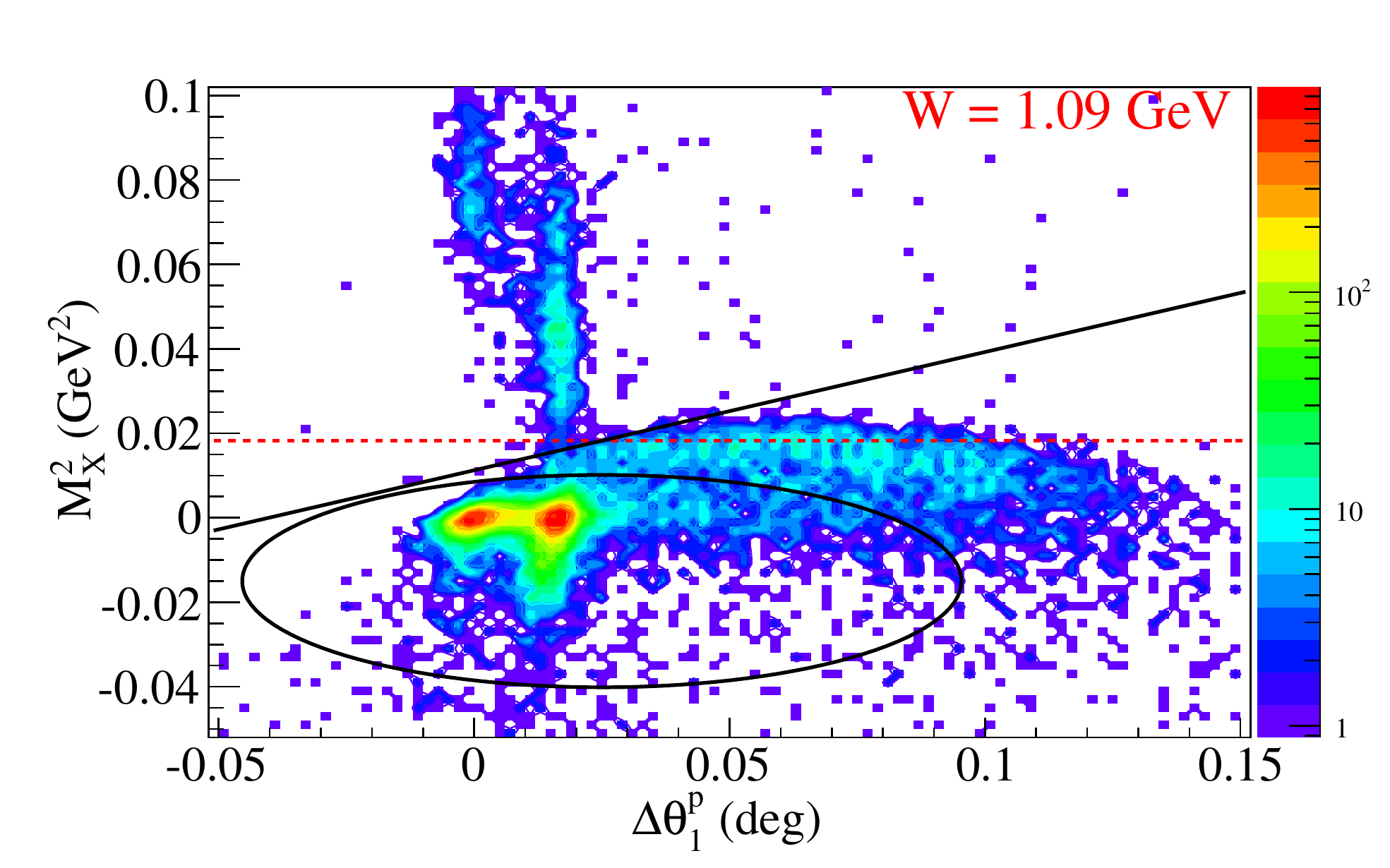}
	  \label{fig:fig8a}
	}
	\subfigure[]{
	  \includegraphics[width=8cm]{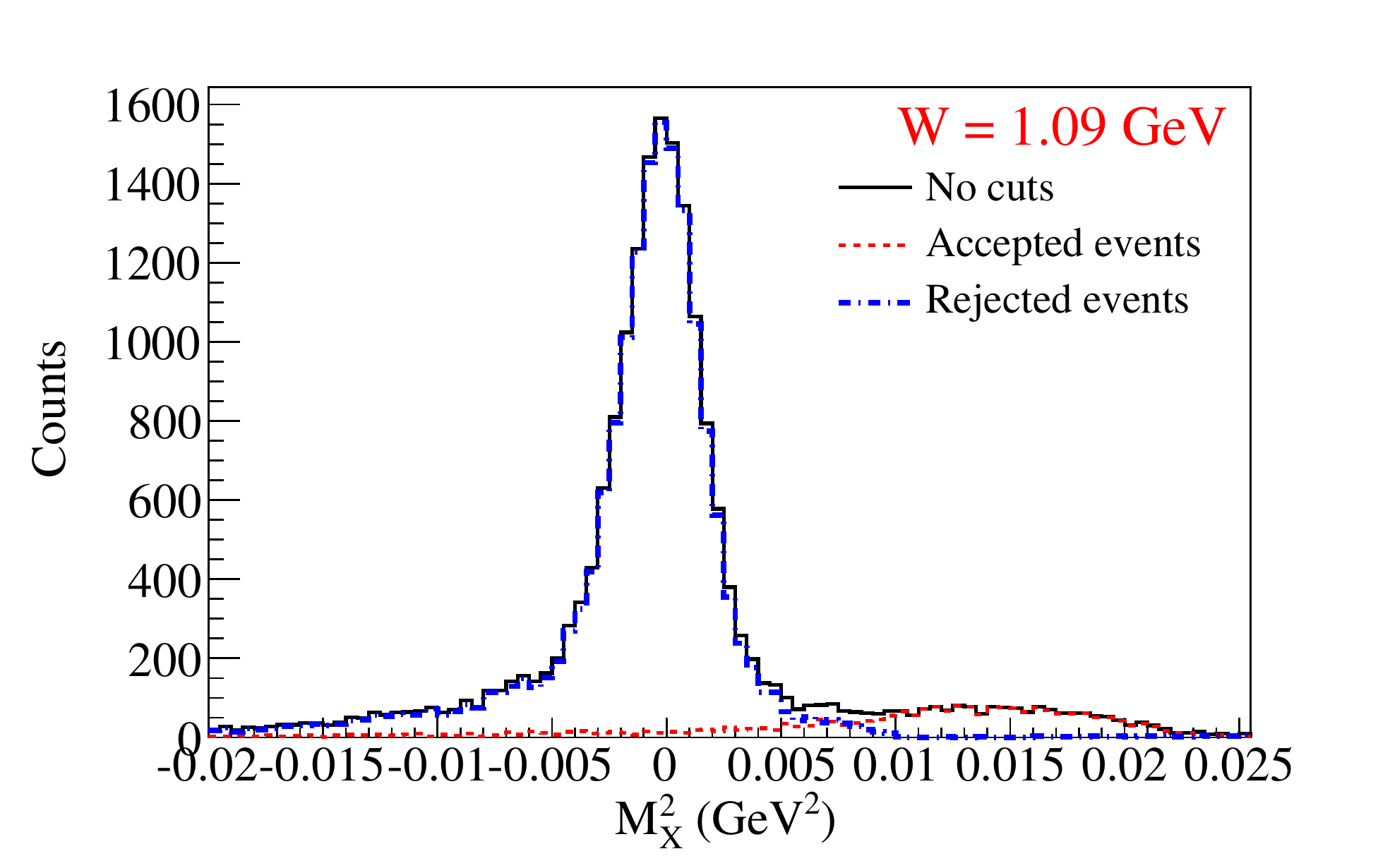}
	  \label{fig:fig8b}
	}
	\caption{(Color online) \subref{fig:fig8a} $M_X^2$ vs $\Delta\theta_1^p$ for
	$W = 1.09 \pm 0.01$ GeV.
	The red dashed line indicates the expected pion peak position.
	The left red spot centered around zero degrees corresponds to the
	elastic scattering events in which the incident electrons have undergone
	Bethe-Heitler radiation (pre-radiative) and the one on the right to the elastic
	post-radiative events. 
	The events below the linear polynomial and outside the ellipse are 
	selected as pions.
	\subref{fig:fig8b} $M_X^2$ for events with $W = 1.09 \pm 0.01$ GeV. 
	The black solid
	curve shows events prior to any Bethe-Heitler subtraction cuts, the blue dashed-dot curve
	shows events rejected from the cuts, and the red dashed curve shows those events
	that survive the Bethe-Heitler subtraction cuts.}
\end{figure}

Fig.~\ref{fig:fig8a} shows the $M_X^2$ plotted as a function of this
deviation $\Delta\theta_1^p$ for one of the near threshold regions, 
$W = 1.09 \pm 0.01$ GeV. In the plot, we see two red spots along 
$M_X^2 = 0 \, \textrm{GeV}^2$. The one on the left is centered along
$\Delta\theta_1^p = 0$ deg corresponding to the pre-radiated photon
events. The other corresponds to the post-radiated photon events. Additionally,
these radiative events are also present in the positive $M_X^2$. These are
the radiative events that we need to isolate from the pion signal as indicated
by the red dashed line in the plot. An ellipse and a linear polynomial are used to reject 
these events. These cuts are parameterized as a function of $W$. 
The result of these cuts is seen in
Fig.~\ref{fig:fig8b} with the accepted events after the cut shown in red (dashed 
curve) as our pions and the rejected events in blue (dashed-dot curve). 

\begin{figure}
  \centering
  \includegraphics[width=8.5cm]{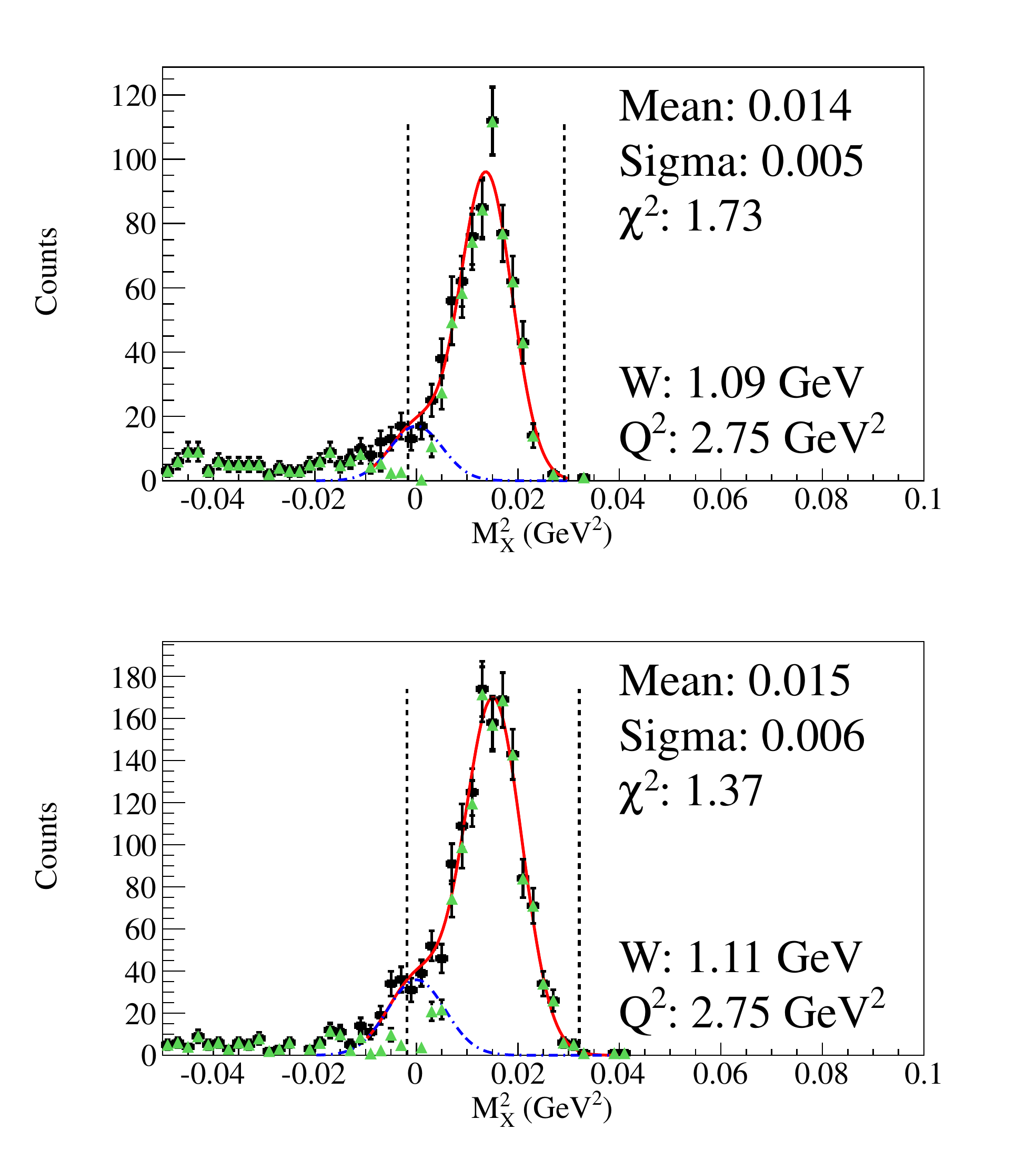}
  \caption{\label{fig:fig9} (Color online)
  An example of the $M_X^2(ep)$ distribution with a 
  double Gaussian fit after applying the
  elliptical cuts (black circles) of Fig.~\ref{fig:fig8a} and after 
  residual Bethe-Heitler and other contamination
  subtractions (green triangles) for $Q^2 = 2.75 \pm 0.25$ 
  GeV$^2$ and $W = 1.09 \pm 0.01$ GeV (top)
  and $W = 1.11 \pm 0.01$ GeV (bottom) integrated over all $\phi^*_\pi$ and
  $\cos\theta^*_\pi$. The black dashed lines indicate the $\pm 3\sigma$ cuts applied to select
  the pions. The $\chi^2$ is the goodness of fit per degree of freedom. See Sec.~\ref{sec:bhsub} for details.}
\end{figure}

After the Bethe-Heitler subtraction cuts are applied, the pions are selected by
making a $\pm 3\sigma$ cut on $M_X^2$ from the mean position of the 
distribution. An example of the distributions and fit are shown in Fig.~\ref{fig:fig9}.  
The $M_X^2$ distributions (black circles) are fit with two Gaussians. The blue (dashed-dot) curve is 
an estimate of the remaining Bethe-Heitler background in the $M_X^2$ distribution, which
was not eliminated by the elliptical cuts of Fig.~\ref{fig:fig8a}.
This was subtracted to yield the green (triangle) points. A systematic uncertainty
of $\pm8\%$ is associated with this background subtraction procedure, which is
detailed in Sec.~\ref{sec:sys}.

\section{Simulations}

To determine the cross section, a Monte Carlo simulation study is required,
including a physics event generator and the detector geometry. 
Events are generated using the MAID2007 unitary isobar model (UIM) 
\cite{DrechselMAID2007}, 
which uses a phenomenological fit to
previous photo- and electroproduction data. Nucleon resonances are described
using Breit-Wigner forms and the non-resonant backgrounds are modeled from Born terms and 
$t$-channel vector-meson exchange. To describe the threshold behavior, Born
terms were included with mixed pseudovector-pseudoscalar  
$\pi NN$ coupling \cite{DrechselMAID2007}.
While the pion electroproduction world-data in the resonance region goes up to
$Q^2 \sim 7$ GeV$^2$ \cite{villano2009} for $W > 1.11$ GeV, there are no data near 
threshold for $Q^2 > 2$ GeV$^2$ and $W < 1.11$ GeV (the kinematics of this work).
Thus, cross sections for the kinematics of this work are described
by extrapolations of the fits to the existing data in the MAID2007 model.

Events are generated to cover the entire kinematic range described in Table 
\ref{sim:kinbin}. About 73 million events are generated for the 2400 
kinematic bins and 6.7 million events were reconstructed after all analysis
cuts. The average resolutions of the kinematic quantities, $W$, $Q^2$, 
$\cos\theta_\pi^*$, and $\phi_\pi^*$ are 0.014 GeV, 0.008 GeV$^2$, 
0.05, and 8 degrees, respectively. These resolutions are obtained by comparing
the generated kinematic quantities with those after reconstruction. 

\begin{table}[t]
	\centering
	\begin{tabular}{l  c  c  c}
		\hline
		\textbf{Variable} & \textbf{Range} & \textbf{Number of Bins} & \textbf{Width}\\
		\hline
		\hline
		$W$ (GeV) & 1.08 : 1.16 & 4 & 0.02\\  
		$Q^2$ (GeV$^2$) & 2.0 : 4.5 & 4 & variable \\ 
		$\cos\theta_\pi^*$ & -1 : 1 & 5 & 0.4 \\ 
		$\phi_\pi^*$ (deg) & 0 : 360 & 6 & 60\\
		\hline
	\end{tabular}
	\caption{\label{sim:kinbin} Kinematic bin selection.}
\end{table}

After the physics events are generated, their passage through the detector
is simulated using the GEANT3 based Monte Carlo (GSIM) program. This program
simulates the geometry of the CLAS detector during the experiment and the
interaction of the particles with the detector material. GSIM models the effects
of multiple scattering of particles in the CLAS detector and geometric
mis-alignments. The information for all interactions with the detectors is
recorded in raw banks, which is used for reconstruction of the tracks. 

The events from GSIM are fed through
a program called the GSIM Post Processor (GPP) to incorporate effects of
tracking resolution and dead wires in the drift chambers, and timing resolutions
of the TOF. 

These events are then processed using the same codes as those events from the experiment to 
reconstruct tracks and higher level information such as 4-momentum, timing, and so on.
The simulated events are analyzed the same way as the experimental data and are used to 
obtain acceptance corrections and radiative corrections for the cross sections calculations.


\section{Corrections}


\subsection{Acceptance Corrections}

Acceptance corrections are applied to the experimental data to obtain the
cross section for each kinematic bin. These corrections describe the geometrical
coverage of the CLAS detector, inefficiencies in hardware and software, and 
resolution effects from track reconstruction. 

\begin{figure}
  \centering
  \includegraphics[width=8.5cm]{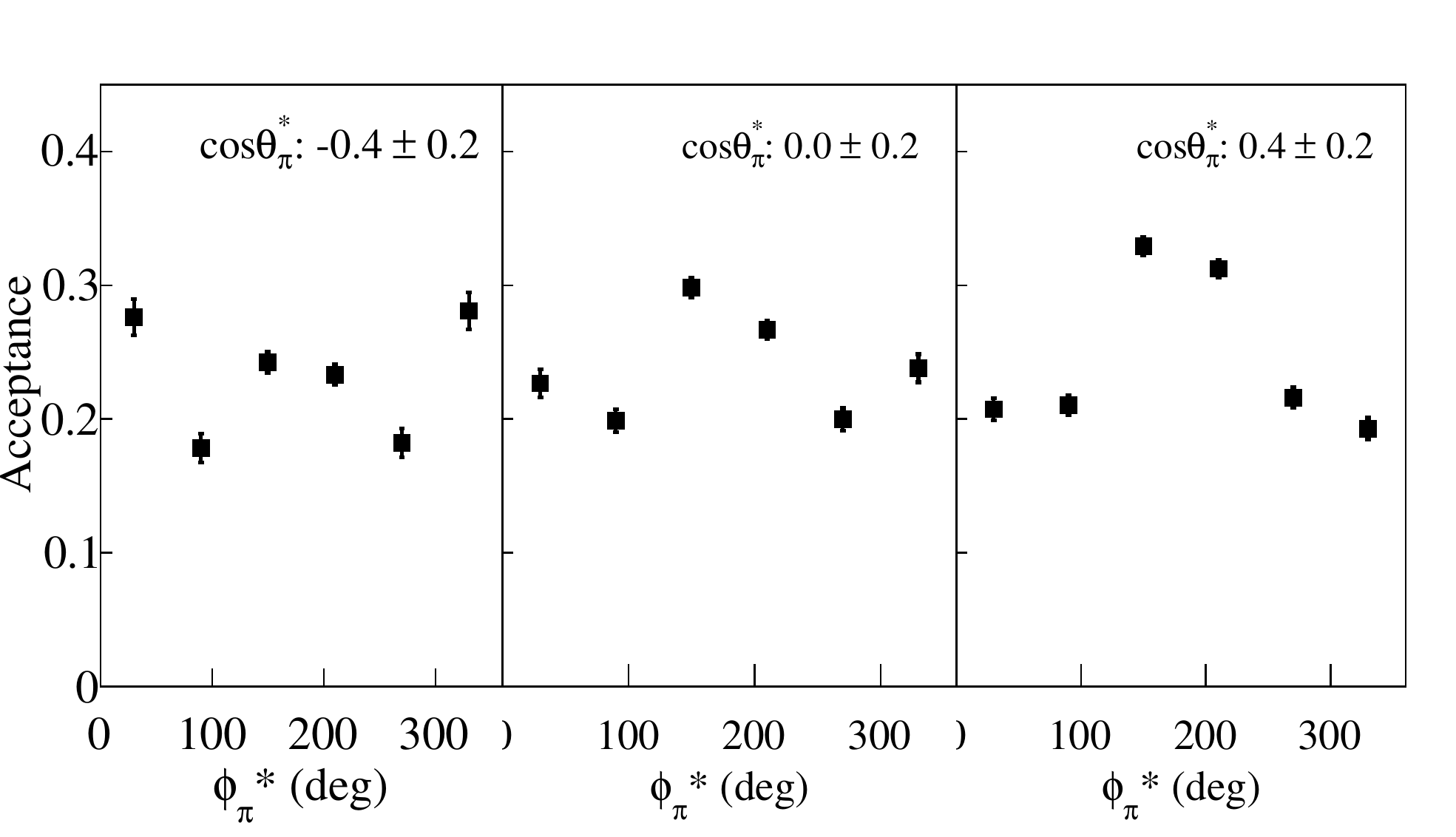}
  \caption{\label{fig:fig10} Acceptance corrections for $W = 1.09$ GeV and $Q^2 = 2.75$ 
  GeV$^2$ as a function of $\phi_\pi^*$. Each subplot shows the correction for a different 
  $\cos\theta_\pi^*$ bin.}
\end{figure}

By comparing the number of events in each kinematic bin from the physics 
generator and the reconstruction process, the acceptance can be obtained as:
\begin{equation}\label{eqn:acc}
	A_i = \frac{N^i_{rec}}{N^i_{gen}},
\end{equation}
where $N^i_{rec}$ corresponds to those events that have gone through the entire
analysis process including track reconstruction and all analysis cuts. $N^i_{gen}$
are those events that were generated. 
Fig.~\ref{fig:fig10} shows the acceptances for a few of the near threshold bins as
a function of $\phi_\pi^*$. 


\subsection{Radiative Corrections}

The radiative correction is obtained using the software package EXCLURAD 
\cite{afanasevradcor} that takes theoretical models as input
to compute the corrections. For this experiment the MAID2007 model, the same model
used to generate Monte Carlo events, is used to determine the radiative corrections.
The radiative corrections are closely related to the acceptance corrections. For
each kinematic bin the differential cross section can be written as:
\begin{equation}
  \sigma = \frac{N_{meas}}{\mathcal{L}\, A}\frac{1}{\delta},
\end{equation}
where $N_{meas}/\mathcal{L}$ is the number of events from the experiment normalized
by the integrated luminosity (with appropriate factors) before acceptance and 
radiative corrections. Also,
$A = N^{RAD}_{rec}/N^{RAD}_{gen}$ is the acceptance correction for the 
bin and $\delta$ is the radiative correction. It should be noted that 
the events for the acceptance correction were generated with a radiated photon
in the final state using the MAID2007 model. 

EXCLURAD uses the same model to obtain the correction
$\delta = N^{RAD'}_{gen}/ N^{NORAD'}_{gen}$, where $N^{NORAD'}_{gen}$ are 
events generated without a radiated photon in the final state. Thus
\begin{equation}
	\sigma = \frac{N_{meas}}{\mathcal{L}} \left(\frac{N^{RAD}_{gen}}{N^{RAD}_{rec}} \right) \times
	\left(\frac{N^{NORAD'}_{gen}}{N^{RAD'}_{gen}} \right).
\end{equation}
The details of the radiative correction procedure are described in 
Ref.~\cite{kijunprc2008}.

\begin{figure}
  \centering
  \includegraphics[width=8cm]{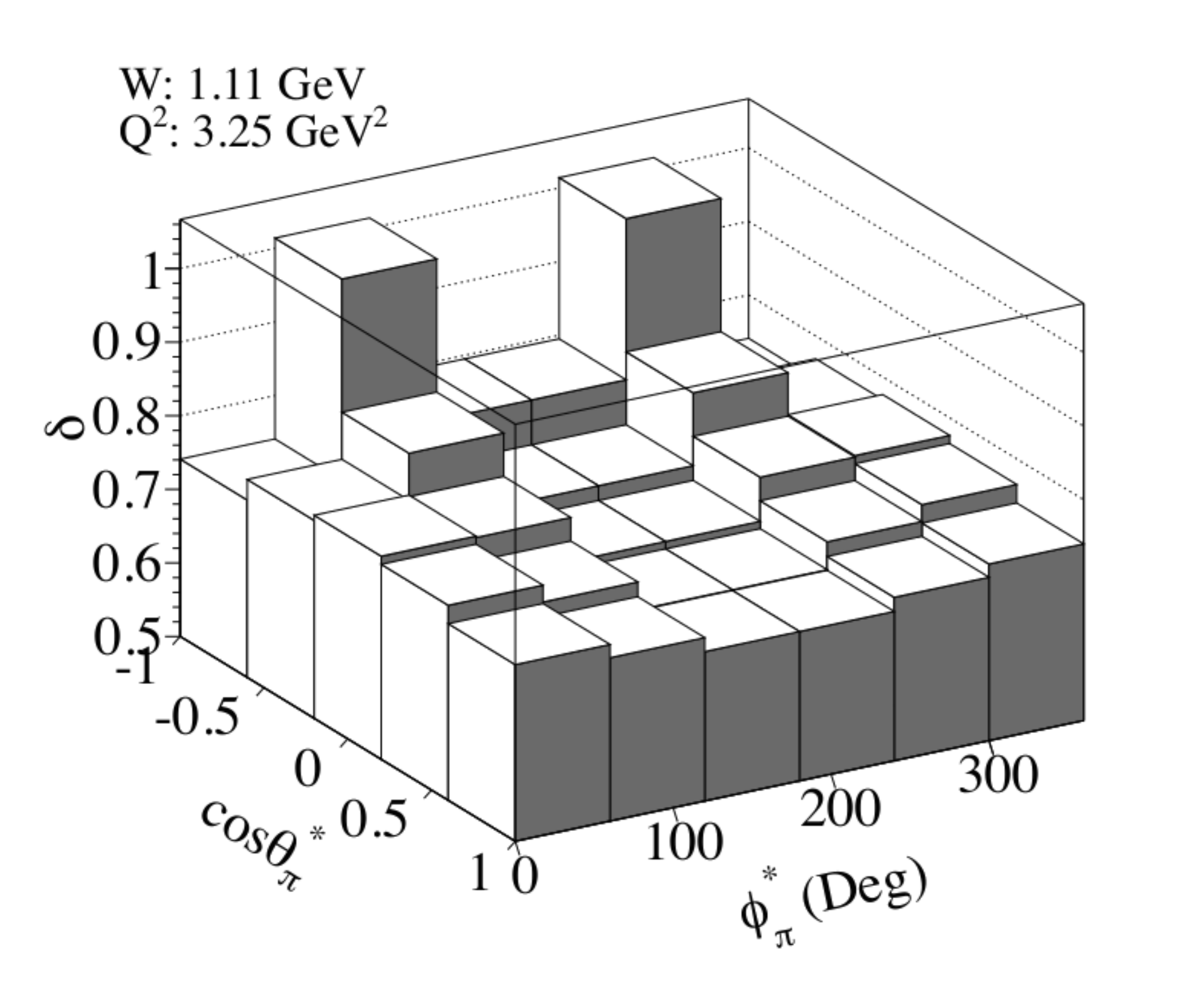}
  \caption{\label{fig:fig11} The radiative corrections for $W = 1.11$ GeV and
  $Q^2 = 3.25$ GeV$^2$ as a function of $\cos\theta_\pi^*$ and $\phi^*_\pi$ obtained
  from EXCLURAD using the MAID2007 model.}
\end{figure}

Fig.~\ref{fig:fig11} shows the radiative corrections calculated for one of the
kinematic bins as a function of the pion angles in the c.m.~system. 
One can observe
that the corrections have a $\phi^*_\pi$ dependence. This is because the
bremsstrahlung process only occurs near the leptonic plane, $i.e.$, 
at angles near 0 or 180 degrees with respect to the hadronic
plane. Also, one can notice that the correction increases with 
$\cos\theta_\pi^* \to -1$. This is because the cross section is expected to 
approach zero at backwards angles and that is the region where the Bethe-Heitler
events dominate. The average
radiative correction over all kinematic bins is $\sim 25\%$.


\subsection{Other Corrections}

Two other corrections were applied to the cross section. One of them involves
estimating the fraction of the events originating from the target cell walls and
the other is an empirical overall normalization factor.

To estimate the level of contamination from the target cell walls, events
collected during the empty-target run period of the experiment
are analyzed using the same process as those for the production run 
period. Only those events that fall within
the target wall region for the empty target should be considered for the source
of contamination. This is because even though there was no liquid hydrogen in the 
target, it was still filled with cold hydrogen gas. So, for this estimation only
events within $\pm 0.5$ cm of the target wall region are selected. The correction
is then calculated by taking the ratio of events within this target region 
from the empty target runs to
those from the production run normalized to the total charge, $\rho$, collected during the
run periods,
\begin{equation}
	R = \frac{N_\mathrm{empty\, target}}{N_\mathrm{production}} 
	\frac{\rho_\mathrm{production}}{\rho_\mathrm{empty\,target}}.
\end{equation}
The average contamination is approximately $1\%-1.9\%$ depending on the $W$ 
kinematic bin. This ratio is then applied as a correction factor to the measured
cross section $\sigma = \sigma_{meas} (1 - R)$. Here, $\sigma$ is the corrected 
cross section and $\sigma_{meas}$ is the measured cross section for a particular bin in $W$.

The second correction (the empirical overall normalization factor) comes from comparing 
the measured 
$ep \to ep$ elastic and the $ep \to ep\pi^0$ cross sections in the $\Delta(1232)$ 
resonance region ($W = 1.23$ GeV) to previously measured values 
\cite{bostedff,ungaroprl,aznauryan,DrechselMAID2007}. 
The measured elastic scattering cross section from this experimental data 
were compared to the known cross section values \cite{bostedff} where both the electron 
and the proton were detected in the final state. A deviation of $\sim 11\%$ from the
known cross section values is observed. 

This deviation of $\sim 11\%$ from the known elastic electron-proton scattering 
cross section includes the inefficiencies associated with the
proton detection in CLAS \cite{clas,ivan2012}.

\begin{figure}
 \center
 \includegraphics[width=8.5cm]{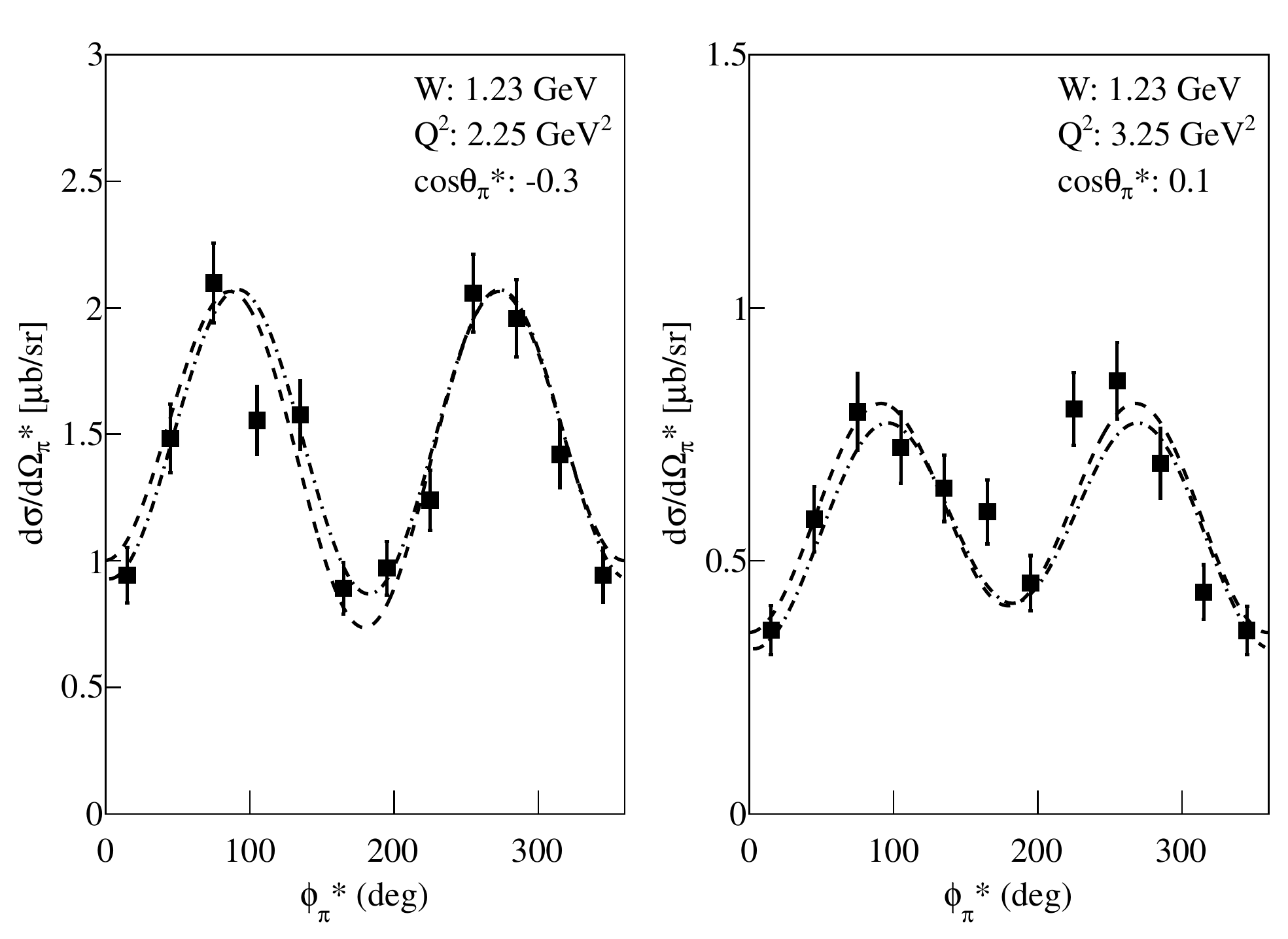}
 \caption{\label{fig:fig12} The differential cross section $ep \to ep\pi^0$ for
 the $\Delta(1232)$ resonance region, $W = 1.23 \pm 0.01$ GeV, for typical
 kinematic bins. The squares are the measured cross sections after applying the 
 normalization correction factor (see text for details). The
 dashed curves are from Ref.~\cite{aznauryan} and the dashed-dot curves are from the
 MAID2007 model. The corrected values agree with the two curves to within 
 $5\%$ on average.}
\end{figure}

To account for this discrepancy, an overall
normalization factor of $R_{elastic} = 0.89$ is applied to the 
$ep \to ep\pi^0$ differential cross section for every kinematic bin. An associated
 systematic uncertainty of $\pm5\%$ is applied. After this correction is
applied, the measured $ep \to ep\pi^0$ cross sections for the $\Delta(1232)$ resonance 
region, $W = 1.23 \pm 0.01$ GeV, are in agreement with previous measurements 
\cite{ungaroprl,aznauryan,DrechselMAID2007} to within $5\%$ on average. 
Fig.~\ref{fig:fig12} shows the result of this correction for a few kinematic
bins in the $\Delta(1232)$ resonance region. 

Since the
threshold region of interest for this experiment is sandwiched between the 
elastic and the $\Delta(1232)$ resonance region and the results in these two regions are
consistent with previous measurements after applying this overall normalization factor,
we believe this procedure is justified. This correction to the cross section also 
includes any detector inefficiencies and, as such, these inefficiencies will not be
accounted for separately. 


\section{Systematic Studies}\label{sec:sys}

To determine the systematic uncertainties in the analysis, the
parameters of the likely sources of those uncertainties are varied within
reasonable bounds and the sensitivity of the final result is checked
against this variation. A summary of the systematic uncertainties averaged 
over the kinematic bins of interest is shown in Table~\ref{tbl:syserr}. 

The electron and proton identification cuts, the electron fiducial cuts,
the vertex cuts and the target cell correction cuts provide small 
contributions to the overall systematic uncertainties.

The electron EC sampling fraction cuts were varied 
from $3\sigma$ to $3.5\sigma$
and the extracted structure functions changed by about $0.4\%$ on average. 
The parameters for the electron fiducial cuts were similarly varied by about
$10\%$ and the structure functions changed
by about $1\%$ on average. As such, a systematic uncertainty of $0.4\%$ and
$1\%$ was assigned to these sources.

The $\Delta t$ cuts to select the protons were varied from $3.5\sigma$ to
$4\sigma$ and a variation of about $1.1\%$ on average was observed on
the extracted structure functions, which was assigned as the systematic
uncertainty associated with this source. The variations in the fiducial cuts for 
the proton had a negligible effect on the structure functions. 

The vertex cuts were reduced by $5\%$ and a variation of about $0.1\%$
on average was observed on the extracted structure functions. So, a systematic
uncertainty of $0.1\%$ was assigned to this source.
The structure functions are compared before and after applying the target cell
corrections. A variation of about $1\%$ is observed and this value was
assigned as a source of systematic uncertainty.  

The major sources of systematic
uncertainty are the Bethe-Heitler background subtraction, the missing mass squared 
cut to select the neutral pions, the elastic normalization corrections and the 
model dependence of the acceptance and radiative corrections. 

There are residual Bethe-Heitler events that escape the elliptical Bethe-Heitler
cuts. These events peak at $M_X^2 = 0$, which have to be included in the overall
fit. A Gaussian distribution was assumed for both the $\pi^0$ and the remaining
Bethe-Heitler events. The pions are modeled by a Gaussian distribution near the
expected pion mass and the Bethe-Heitler events are modeled by a Gaussian 
whose peak is at $M_X^2 = 0$. This accounts for much of the tail in 
Figs.~\ref{fig:fig8b} and \ref{fig:fig9}. The resolution for $M_X^2$ for the 
Bethe-Heitler and the pion distributions is expected to be similar because
of the same kinematics of the detected electron and the proton. The 
Gaussian fit for the Bethe-Heitler is obtained, which is then subtracted to yield
the pions. 

To see the effect of the background subtraction, the structure functions were 
compared with and without the application of the Bethe-Heitler background 
subtraction cuts. 
The structure functions changed by about $8\%$ on average and this was used as a
systematic uncertainty for this procedure. 

The missing mass squared cut was varied from $3\sigma$ to $4\sigma$ and this 
resulted in a change of about $3\%$ on average in the extracted structure functions. 

The systematic uncertainty on the elastic normalization correction of $\pm5\%$
was obtained by looking at the difference between the extracted structure
functions before and after applying the correction factor to the data. The
structure functions varied by about $5\%$ on average.

Additionally, a $\pm4\%$ systematic uncertainty is assigned on the model dependence 
of the acceptance and radiative corrections based on previous analyses 
\cite{ungaroprl,kijunprc2008,kijun2012}.

The total average systematic uncertainty, obtained by adding the individual 
contributions in quadrature is 10.8\%.

\begin{table}
	\centering
	\begin{tabular}{l c}
		\hline
		\textbf{Source} & \textbf{Estimate \%}\\
		\hline
		\hline
		$e^-$ EC sampling fraction cuts         & 0.4 \\
		$e^-$ fiducial cuts                     & 1 \\
		$p$ $\Delta t$ cuts                     & 1.1 \\		         
        Vertex cuts                             & 0.1 \\ 
		Background subtraction cuts             & 8 \\
        $\pi^0$ $M_X^2$ cut                     & 3 \\
        Target cell correction                  & 1\\
        Elastic normalization correction        & 5 \\
        Acceptance and radiative correction     & 4 \\
        \hline
        Total                                   & 10.8 \\
		\hline
	\end{tabular}
	\caption{\label{tbl:syserr} The average systematic uncertainties for the differential 
	cross sections from various sources and the corresponding criteria. The final
	quoted systematic uncertainty, obtained by adding the different systematic
	uncertainties from each source in quadrature, is about 10.8\%.}
\end{table}


\section{Differential Cross Sections and Structure Functions\label{sec:cross_sf}}

\begin{figure*}[htpb]
  \centering
  \includegraphics[width=8cm]{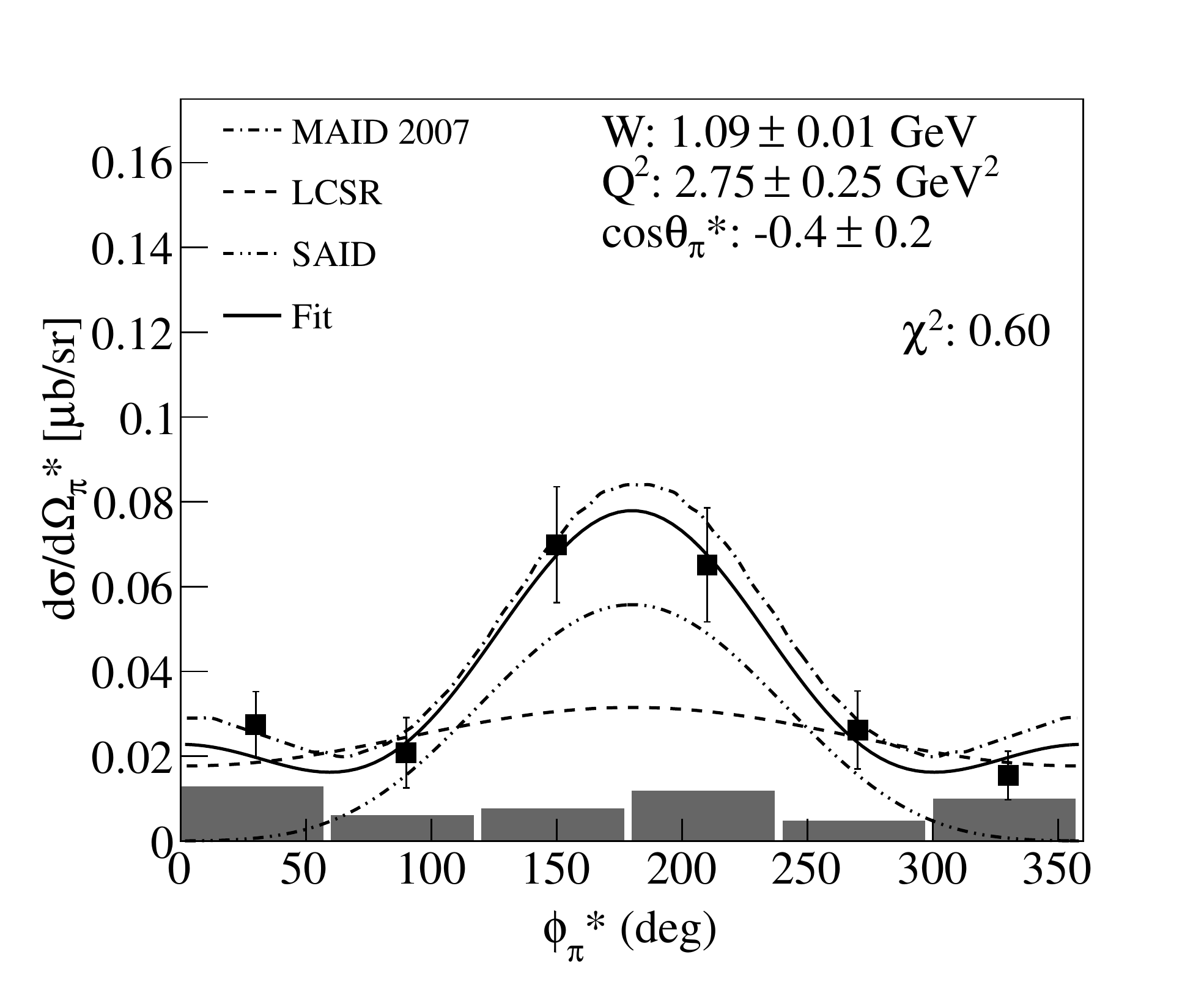}
  \includegraphics[width=8cm]{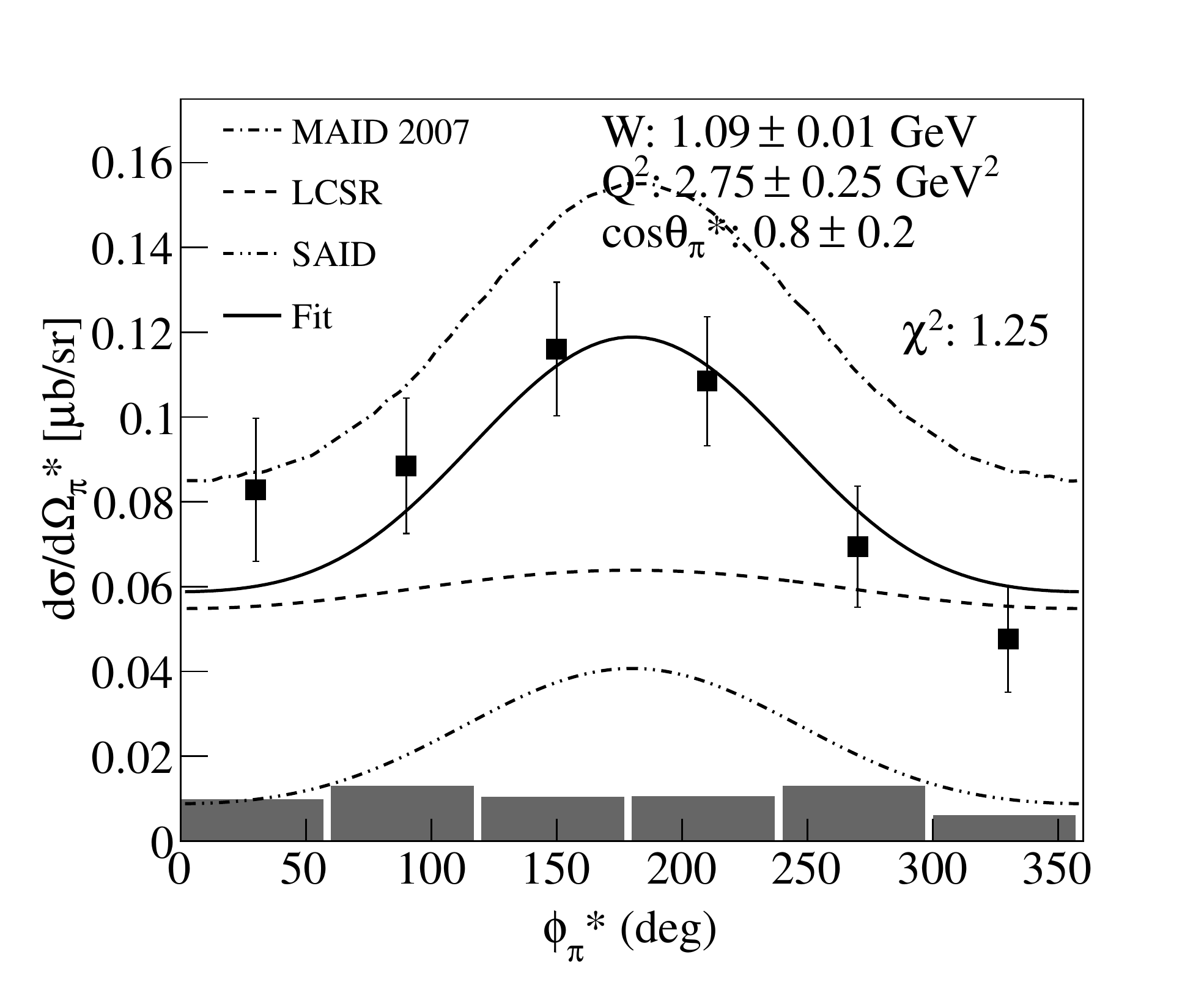}\\
  \includegraphics[width=8cm]{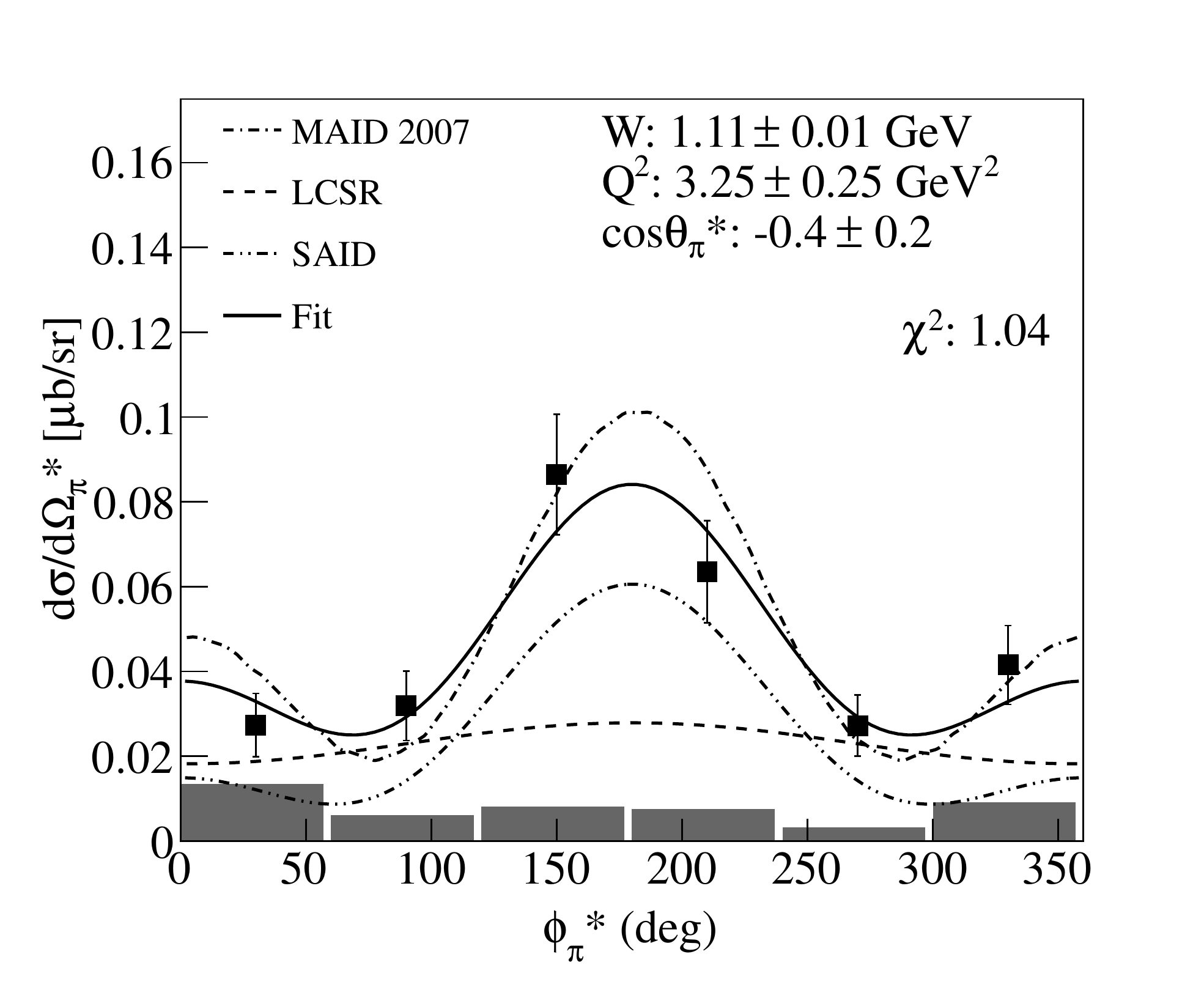}
  \includegraphics[width=8cm]{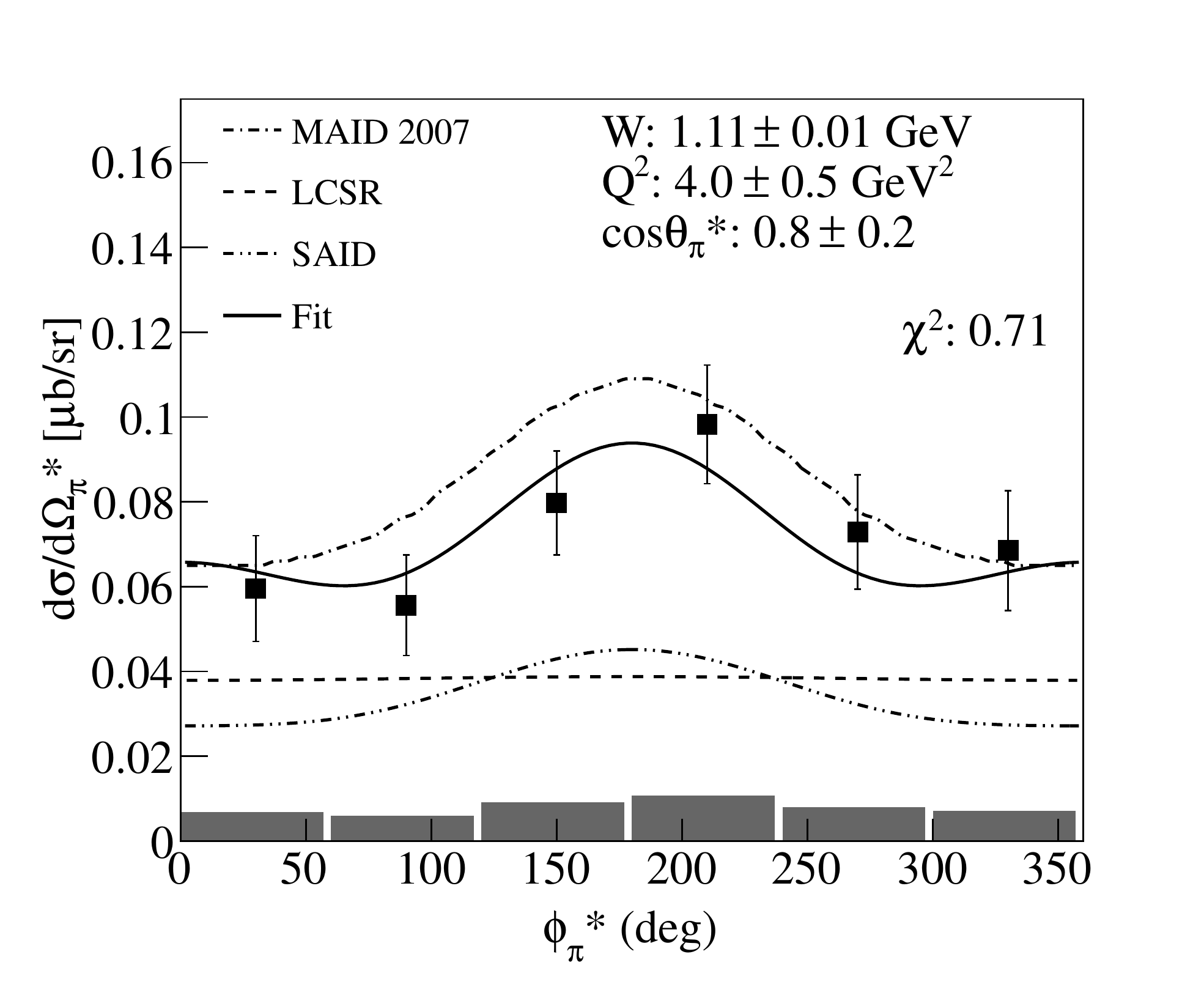}
  \caption{\label{fig:fig13} The differential cross sections in $\mu$b/sr for
  a few kinematic bins near threshold as a function of $\phi_\pi^*$. Experimental
  points (squares) are shown with statistical uncertainties only. The size of the estimated
  systematic uncertainties is shown in gray boxes below. The predictions from 
  LCSR, MAID2007 and SAID are shown as dashed, dashed-dotted and dashed-double-dotted curves, 
  respectively. The fit to the distributions is shown as a solid curve. See 
  Sec.~\ref{sec:cross_sf} for details.}
\end{figure*}

The kinematic coverage of the experiment spans over $W$ from $1.08$ to $1.16$ GeV and 
$Q^2$ from $2$ to $4.5$ GeV$^2$. 
The reduced differential cross section for the reaction is computed for each 
kinematic bin. The cross sections are reported at the center of each kinematic
bin. Fig.~\ref{fig:fig13} shows the differential cross section for some
of the kinematic bins near threshold as a function of $\phi_\pi^*$. 
The predictions from LCSR \cite{braunrev}, MAID2007 \cite{DrechselMAID2007} 
and SAID \cite{SAID} are shown for comparison. 

Using Eq.~(\ref{eqn:diffeqsf}), the differential cross section 
is fitted to extract the structure functions
$\sigma_{T} + \varepsilon\sigma_L$, $\sigma_{TT}$ and $\sigma_{LT}$. 
The result of the fit is shown as the solid curve in 
Fig.~\ref{fig:fig13}. The reduced $\chi^2$ for the fit is calculated using 
$\chi^2 = \chi_0^2/\nu$, where $\nu$ is the number of degrees of freedom calculated for 
each $W$, $Q^2$, and $\cos\theta_\pi^*$ bin (\textit{i.e.}, $\nu = 6$ data points 
$- 3$ fit parameters $= 3$), and $\chi_0^2$ is the unnormalized goodness of fit. 
The averaged $\chi^2$ of the fits is 0.9. 

\begin{figure}[htpb]
  \centering
  \includegraphics[width=8.5cm]{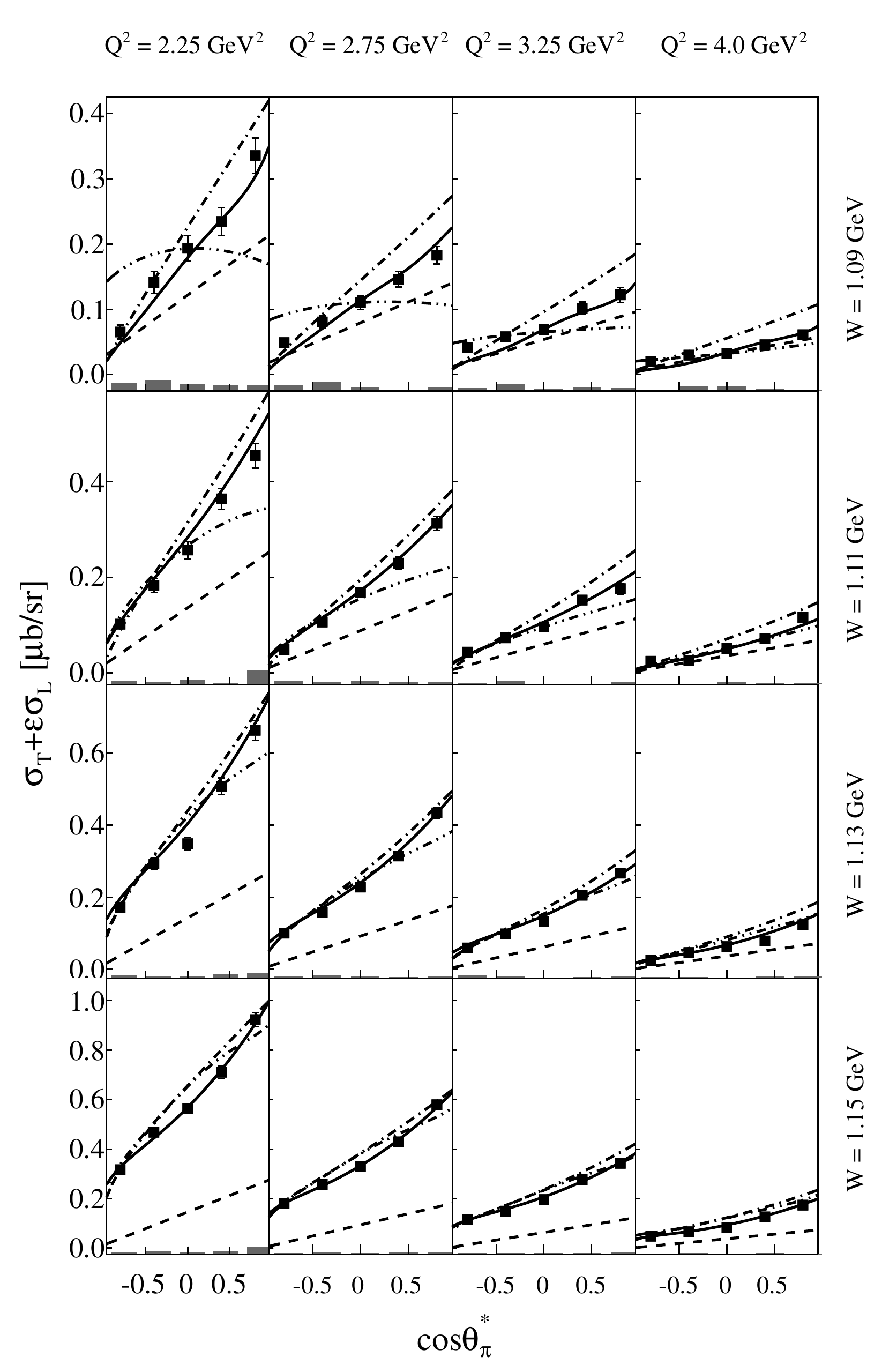}
  \caption{\label{fig:fig14} The structure function $\sigma_T + \varepsilon\sigma_L$
  as a function of $\cos\theta_\pi^*$ in $\mu$b/sr for $W=1.08-1.16$ GeV and 
  $Q^2=2.0-4.5$ GeV$^2$. Predictions from LCSR that include only $s$-wave 
  contribution (dashed), MAID2007 (dashed-dot), and
  SAID (dashed-double-dot) are shown. The error bars represent statistical uncertainties 
  only and the
  estimated systematic uncertainties are shown as gray boxes.
  The solid curve corresponds to the results obtained from the 
  fit to the cross sections (see Sec.~\ref{sec:multgen} for details). The values of
  $Q^2$ (on top of the panels) and $W$ (on the right side of the panels) are the central
  values of the bins.
  }
\end{figure}

\begin{figure}[htpb]
  \centering
  \includegraphics[width=8.5cm]{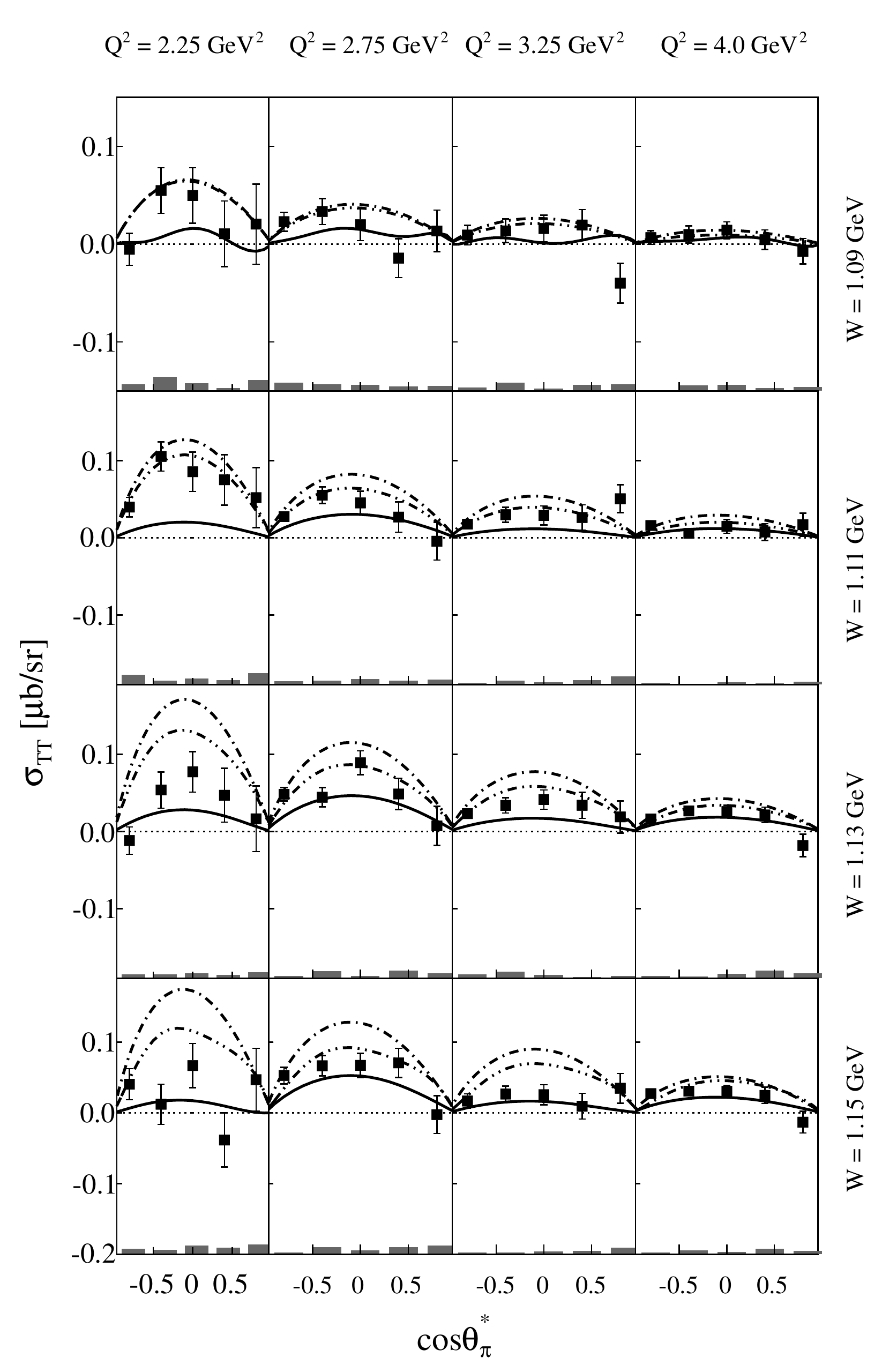}
  \caption{\label{fig:fig15} The structure function $\sigma_{TT}$
  as a function of $\cos\theta_\pi^*$ in $\mu$b/sr for $W=1.08-1.16$ GeV and 
  $Q^2=2.0-4.5$ GeV$^2$. Predictions from MAID2007 
  (dashed-dot) and SAID (dashed-double-dot) are shown. The LCSR predictions do not include any 
  $\sigma_{TT}$
  contributions, so they are not shown. The error bars represent statistical
  uncertainties only and 
  the estimated systematic uncertainties are shown as gray boxes.
  The solid curve corresponds to the results obtained from the 
  fit to the cross sections (see Sec.~\ref{sec:multgen} for details).
  The values of $Q^2$ (on top of the panels) and $W$ (on the right side of the panels) are 
  the central values of the bins. The horizontal line at zero serves as a visual aid.
  }
\end{figure}

\begin{figure}[htpb]
  \centering
  \includegraphics[width=8.5cm]{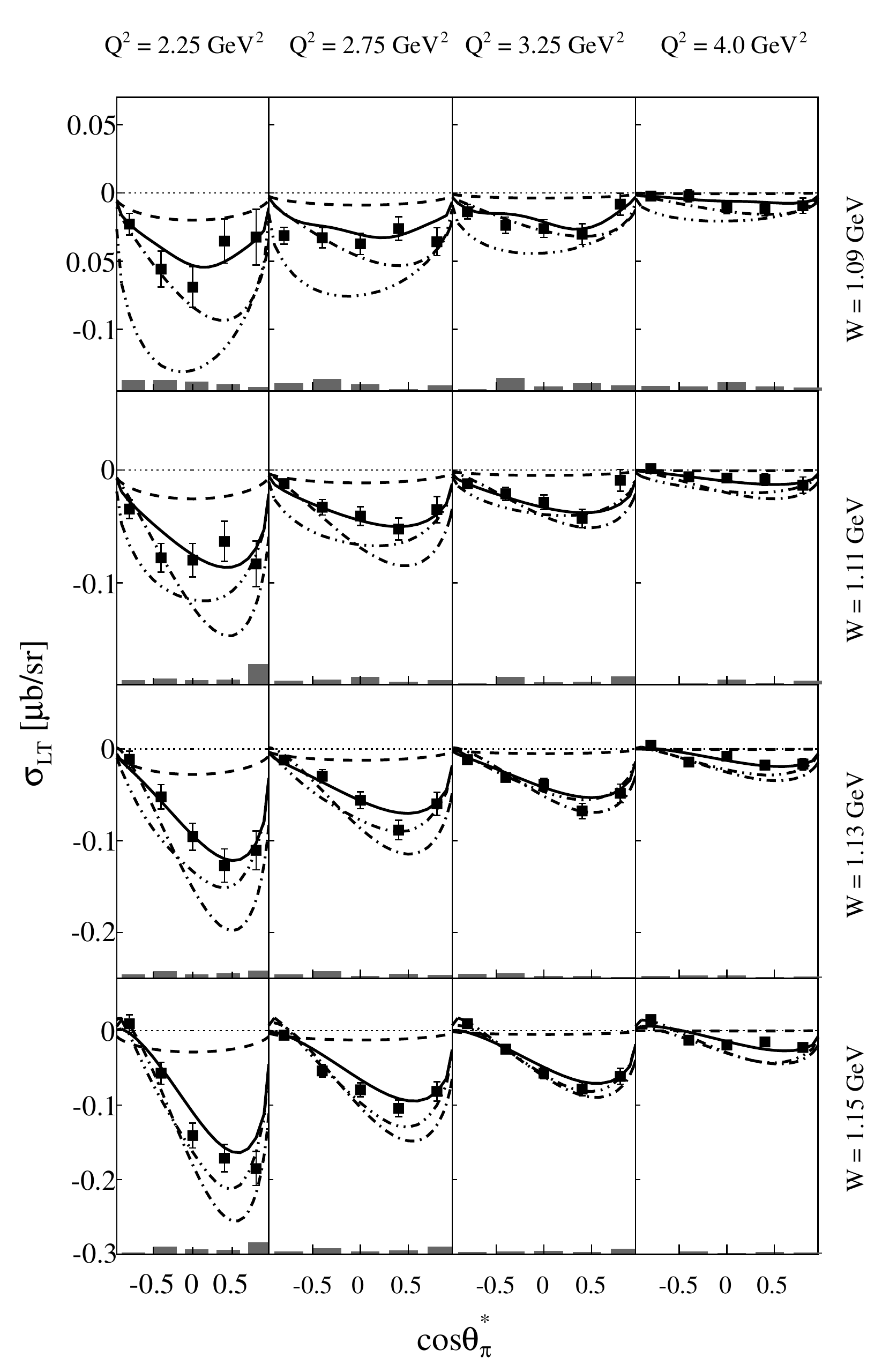}
  \caption{\label{fig:fig16} The structure function $\sigma_{LT}$
  as a function of $\cos\theta_\pi^*$ in $\mu$b/sr for $W=1.08-1.16$ GeV and 
  $Q^2=2.0-4.5$ GeV$^2$. Predictions from LCSR that include only
  $s$-wave contribution (dashed), MAID2007 
  (dashed-dot), and SAID (dashed-double-dot) are shown. The error bars represent 
  statistical uncertainties only and the
  estimated systematic uncertainties are shown as gray boxes.
  The solid curve corresponds to the results obtained from the 
  fit to the cross sections (see Sec.~\ref{sec:multgen} for details). The values of
  $Q^2$ (on top of the panels) and $W$ (on the right side of the panels) are the central
  values of the bins.
  The horizontal line at zero serves as a visual aid. 
  }
\end{figure}

The extracted structure functions $\sigma_{T}+\varepsilon\sigma_{L}$, 
$\sigma_{TT}$ and $\sigma_{LT}$ are shown in Figs.~\ref{fig:fig14}, \ref{fig:fig15}
and \ref{fig:fig16}, respectively, as a function of $\cos\theta_\pi^*$ for $W=1.08-1.16$
GeV and $Q^2 = 2.0-4.5$ GeV$^2$. The data points are shown with statistical error bars
only and the size of the systematic errors is shown as the gray boxes. 
Predictions from LCSR, MAID2007, and SAID are
also included for $\sigma_{T}+\varepsilon\sigma_{L}$ and $\sigma_{LT}$. Since 
the LCSR does not include any $\sigma_{TT}$ contributions in the calculations, 
they are not shown.

The structure function $\sigma_{T}+\varepsilon\sigma_{L}$ (Fig.~\ref{fig:fig14}) is 
generally in good agreement with the MAID2007 predictions but there is some discrepancy for 
$W = 1.09$ GeV at high $\cos\theta_\pi^*$. This discrepancy is reduced for higher 
$W$ bins. The results disagree with the LCSR predictions, especially for
those bins away from threshold ($W > 1.09$ GeV). This disagreement is also
apparent for low $Q^2$ bins. As one moves closer to threshold and at high $Q^2$,
the agreement is quite good, especially at backward angles $\cos\theta_\pi^* \to -1$.
The LCSRs have been calculated and tuned especially for the 
threshold region at high $Q^2$ and thus, there exists a strong disagreement at higher $W$ 
and low $Q^2$ bins. The predictions from SAID strongly disagree for the first
$W$ bin and low $Q^2$ bins, but converge toward the MAID2007 predictions for 
higher $W$ and $Q^2$. 

The structure function $\sigma_{TT}$ (Fig.~\ref{fig:fig15}) 
results are in good agreement with the SAID and MAID2007
predictions for low $W$ and high $Q^2$ but disagree at high $W$ and low $Q^2$ bins.
Most of the values are close to zero for all $W$.
The LCSR predictions assume only $s$-wave contributions 
to the cross section from this structure function. The $d$-wave contribution
to the total cross sections in SAID range from $0$ to $0.001$ 
$\mu$b for the near threshold bins \cite{SAID}. 

The structure function $\sigma_{LT}$ (Fig.~\ref{fig:fig16}) also shows good agreement 
with the MAID2007 and LCSR predictions for high $Q^2$ and low $W$, but there is some 
discrepancy at other kinematics. The SAID prediction has a large disagreement at low $W$ 
and $Q^2$, but the level of agreement at other kinematics is similar to the MAID2007 model.


\section{$S$-Wave Multipoles and Generalized Form Factors\label{sec:multgen}}

\begin{figure}
  \centering
  \subfigure[]{
    \includegraphics[width=8.25cm]{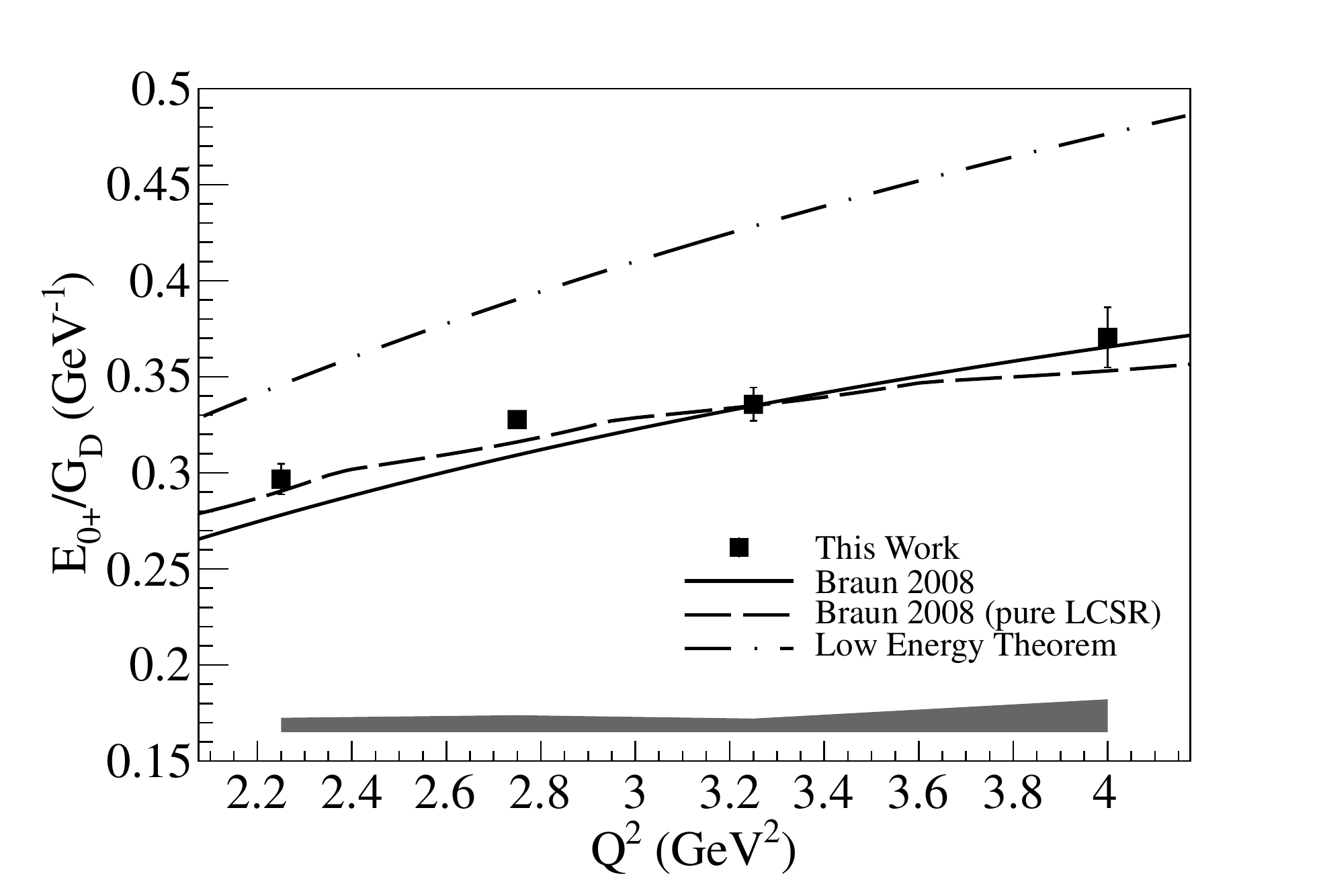}
    \label{fig:fig17a}
  }
  \subfigure[]{
    \includegraphics[width=8.25cm]{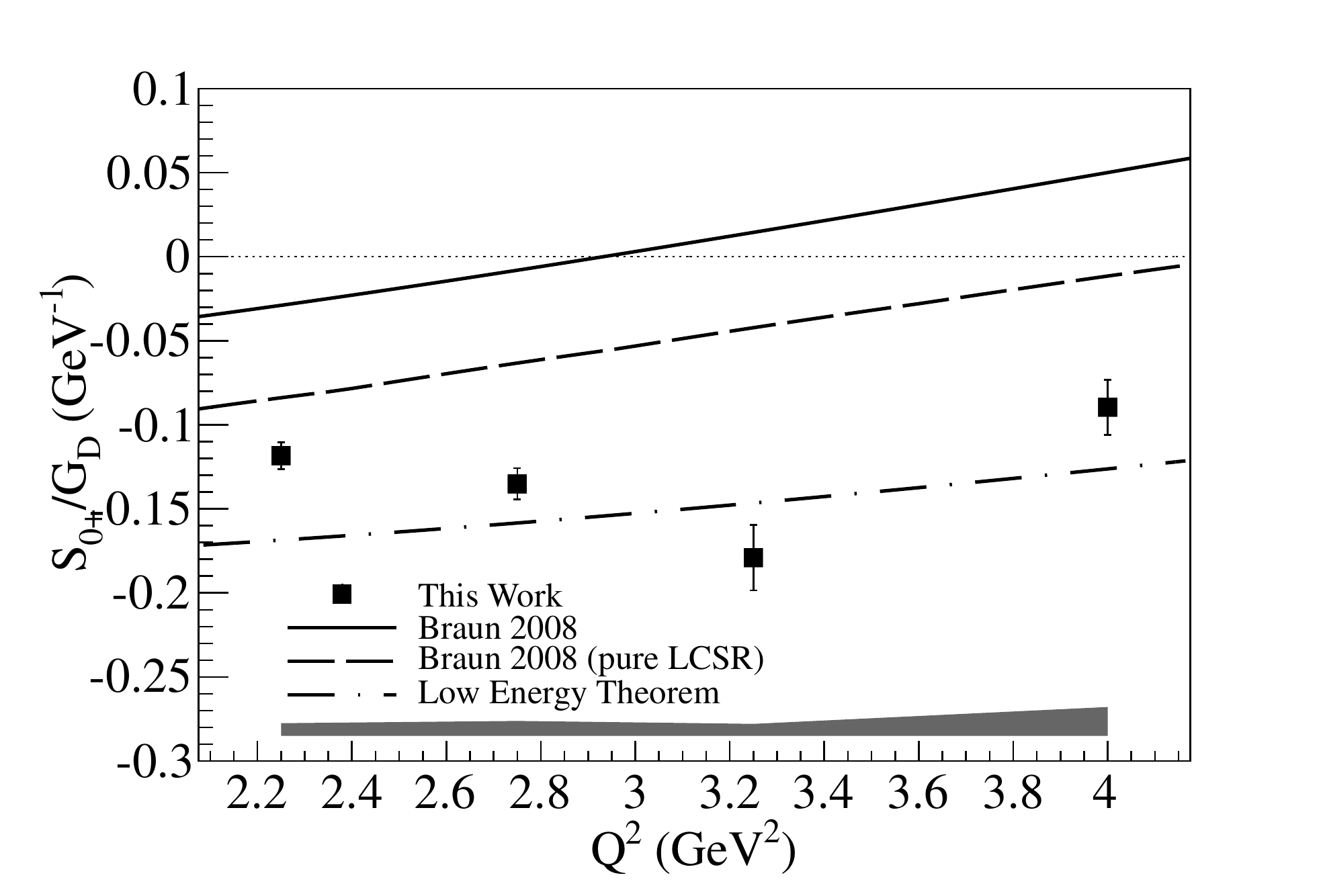}
    \label{fig:fig17b}
  }
  \caption{The $s$-wave multipoles \subref{fig:fig17a} $E_{0+}$ and 
    \subref{fig:fig17b} $S_{0+}$ normalized to the dipole formula $G_D$ are plotted
    as a function of $Q^2$. The error bars include statistical and systematic
    uncertainties added in quadrature. The size of the estimated systematic 
    uncertainties are shown in the bottom. The LCSR based model predictions
    and the LET predictions are also shown as curves.
    The horizontal line at zero serves as a visual aid. 
    \label{fig:mult}}
\end{figure}

\begin{figure}
  \centering
  \subfigure[]{
    \includegraphics[width=8.25cm]{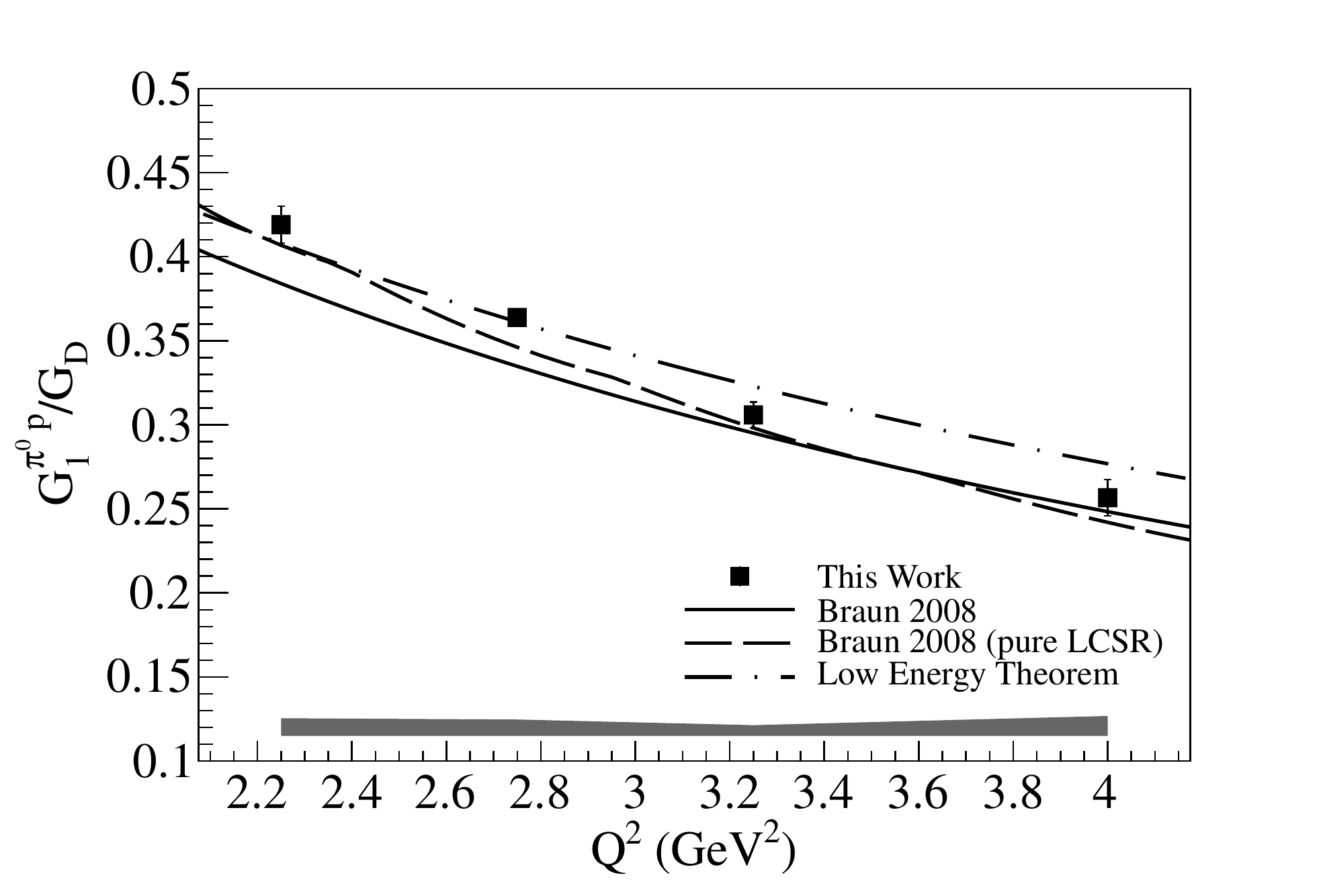}
    \label{fig:fig18a}
  }
  \subfigure[]{
    \includegraphics[width=8.25cm]{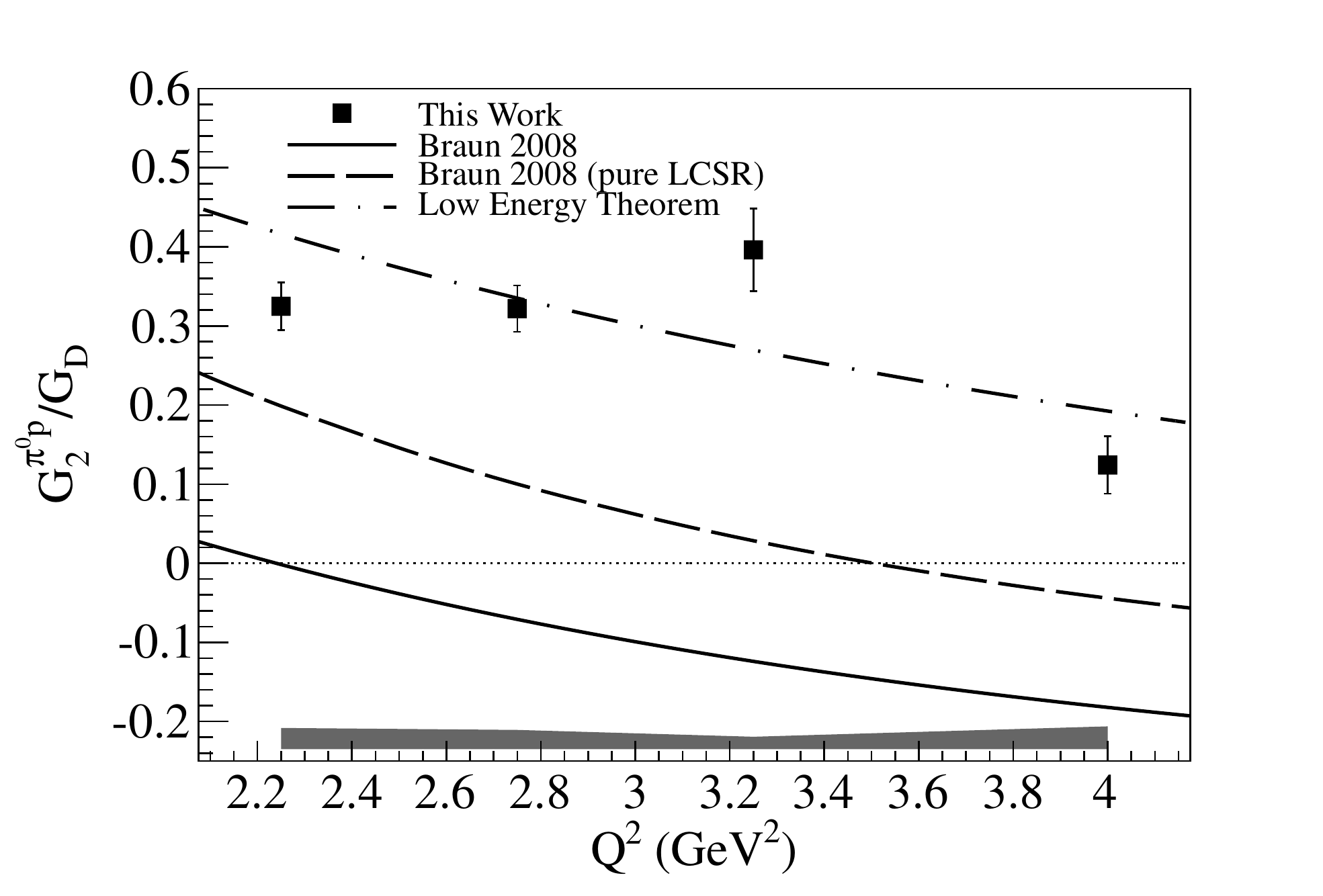}
    \label{fig:fig18b}
  }
  \caption{The generalized form factors \subref{fig:fig18a} $G_{1}^{\pi^0p}$ and 
    \subref{fig:fig18b} $G_{2}^{\pi^0p}$ normalized to the dipole formula $G_D$ are plotted
    as a function of $Q^2$. The error bars include statistical and systematic
    uncertainties added in quadrature. The size of the estimated systematic 
    uncertainties are shown in the bottom. The LCSR based model predictions
    and the LET predictions are also shown as curves.
    The horizontal line at zero serves as a visual aid. 
    \label{fig:g1g2}}
\end{figure}

In order to compare with the calculated generalized form factors of Ref.~\cite{braunrev}, one must 
extract the $s$-wave multipole amplitudes from the measured cross sections. First,
the structure functions are written in terms of the helicity amplitudes
$H_i$. The helicity amplitudes are functions defined by transitions between
eigenstates of the helicities of the nucleon and the virtual photon \cite{amaldi}.
The helicity amplitudes are then expanded in terms of the multipole amplitudes. 

The structure functions are related to the helicity amplitudes
$H_{1,2,...6}(W,Q^2,\cos\theta_\pi^*)$ by:
\begin{eqnarray} 
\sigma_T&=&\frac{1}{2}(|H_1|^2+|H_2|^2+|H_3|^2+|H_4|^2),
\label{eq:eq7}\\
\sigma_L&=&
|H_5|^2+|H_6|^2,
\label{eq:eq8}\\
\sigma_{TT}&=&Re(H_3H_2^*-H_4H_1^*),
\label{eq:eq9}\\
\sigma_{LT}&=&-\frac{1}{\sqrt{2}}
Re[(H_1-H_4)H_5^*+(H_2+H_3)H_6^*].
\label{eq:eq10}
\end{eqnarray}

The analysis of the data is based on the following expansion
of the helicity amplitudes over multipole amplitudes 
(see, for example, \cite{aznauryan2011}): 
\begin{eqnarray} 
H_1&=&\frac{1}{\sqrt{2}}
\sin\theta^*_{\pi}
\cos\frac{\theta^*_{\pi}}{2}
\sum(B_{l+}-B_{(l+1)-})
\nonumber \\
&&[P''_{l}(\cos
\theta^*_{\pi})-P''_{l+1}(\cos
\theta^*_{\pi})],
\label{eq:eq1}\\
H_2&=&\sqrt{2}
\cos\frac{\theta^*_{\pi}}{2}
\sum(A_{l+}-A_{(l+1)-})
\nonumber \\
&&[P'_{l}(\cos
\theta^*_{\pi})-P'_{l+1}(\cos
\theta^*_{\pi})],
\label{eq:eq2}\\
H_3&=&\frac{1}{\sqrt{2}}
\sin\theta^*_{\pi}
\sin\frac{\theta^*_{\pi}}{2}
\sum(B_{l+}+B_{(l+1)-})
\nonumber \\
&&[P''_{l}(\cos
\theta^*_{\pi})+P''_{l+1}(\cos
\theta^*_{\pi})],
\label{eq:eq3}\\
H_4&=&\sqrt{2}
\sin\frac{\theta^*_{\pi}}{2}
\sum(A_{l+}+A_{(l+1)-})
\nonumber \\
&&[P'_{l}(\cos
\theta^*_{\pi})+P'_{l+1}(\cos
\theta^*_{\pi})],
\label{eq:eq4}\\
H_5&=&
\frac{Q}{|\mathbf{q^*}|}
\cos\frac{\theta^*_{\pi}}{2}
\sum(l+1)(S_{l+}+S_{(l+1)-})
\nonumber \\
&&[P'_{l}(\cos
\theta^*_{\pi})-P'_{l+1}(\cos
\theta^*_{\pi})],
\label{eq:eq5}\\
H_6&=&
\frac{Q}{|\mathbf{q^*}|}
\sin\frac{\theta^*_{\pi}}{2}
\sum(l+1)(S_{l+}-S_{(l+1)-})
\nonumber \\
&&[P'_{l}(\cos
\theta^*_{\pi})+P'_{l+1}(\cos
\theta^*_{\pi})].
\label{eq:eq6}
\end{eqnarray}
Here, $P'_{l,l+1}(\cos\theta_\pi^*)$ and $P''_{l,l+1}(\cos\theta_\pi^*)$ are the first 
and second derivatives of the Legendre polynomials, respectively, and
$\mathbf{q^*}$ is the virtual photon 3-momentum in the c.m.~system. Also, 
\begin{eqnarray}
  A_{l+} & = & \frac{1}{2}\left[(l+2)E_{l+} + lM_{l+}\right],  \\
  B_{l+} & = & E_{l+} - M_{l+}, \\
  A_{(l+1)-} & = & \frac{1}{2}\left[ (l+2)M_{(l+1)-} - lE_{(l+1)-} \right], \\
  B_{(l+1)-} & = & E_{(l+1)-} + M_{(l+1)-}.  
\end{eqnarray}

The strong $\cos\theta^*_{\pi}$-dependence of the structure function 
$\sigma_T+\varepsilon \sigma_L$ and the nonzero values of $\sigma_{LT}$ found
in the experiment (see Figs.~\ref{fig:fig14} and \ref{fig:fig16}) show that 
higher multipole amplitudes should be taken into account in addition to 
the $s$-wave amplitudes $E_{0+}$ and $S_{0+}$ at all $W$. 
Our understanding of the high-wave multipoles, which should be included in 
this analysis, was based on the results of the analysis of CLAS data 
\cite{ungaroprl,kijunprc2008} performed in 
Ref.~\cite{aznauryan} using the unitary isobar model (UIM) and dispersion 
relations (DR). 
These data are on the $\gamma^* p\rightarrow \pi^+ n$ \cite{kijunprc2008}
and $\gamma^* p\rightarrow \pi^0 p$ \cite{ungaroprl} cross sections in a 
similar range of $Q^2$ but in a significantly wider energy range,
which start from $W=$ 1.15 and 1.11 GeV, respectively.
The precision in the present experimental results near threshold is much 
better than the precision in Refs.~\cite{ungaroprl,kijunprc2008}.
However, the results
of their analysis are useful to study the $p$- and $d$-wave 
contributions, which are determined mainly by the $\Delta(1232)P_{33}$, 
$N(1440)P_{11}$, and $N(1520)D_{13}$ resonances.

According to the results of the analysis \cite{aznauryan} 
at $W=1.09$ to $1.15$ GeV, there are large $p$-wave contributions
related to the $\Delta(1232)P_{33}$ and $N(1440)P_{11}$.
The $d$-wave contributions are negligibly small for the following reasons: 
(i) near threshold, the $d$-wave multipole amplitudes are suppressed compared 
to the $p$-wave amplitudes by the additional kinematical factor $p^*_\pi$; 
(ii) at the values of $Q^2$ investigated in this experiment, the contribution of 
the $N(1520)D_{13}$ to the corresponding multipole amplitudes is significantly
smaller than the contributions of the $\Delta(1232)P_{33}$ and $N(1440)P_{11}$ to 
the $p$-wave multipole amplitudes; 
(iii) in contrast with the $\Delta(1232)P_{33}$ 
and $N(1440)P_{11}$, the width of the $N(1520)D_{13}$ is significantly smaller than 
the difference between the mass of the resonance and total energy at the threshold. 
Therefore, in our analysis only multipole amplitudes $E_{0+}$, $S_{0+}$, $M_{1\pm}$,
$S_{1\pm}$, and $E_{1+}$ were included.

The data were fitted simultaneously at $W=$ 1.09, 1.11, 1.13 and 1.15 
GeV with statistical and systematic uncertainties added in quadrature for each 
point. The amplitudes were parametrized according to their
threshold behavior and the results of the analysis in Ref.~\cite{aznauryan}.

Due to the Watson theorem \cite{watson1954}, the imaginary parts of the multipole amplitudes
below the $2\pi$ production threshold are related to their real parts as
$Im\mathcal{M} = Re\mathcal{M} \tan(\delta_{\pi N}^I)$, where $\mathcal{M}$
denotes $E^I_{l\pm}$, $M^I_{l\pm}$ or $S^I_{l\pm}$ amplitudes, and $I$ is
the total isotopic spin of the $\pi N$ system. Near threshold 
$\delta^I_{\pi N} \sim p_{\pi}^{*2l+1}$, and the imaginary parts of
the multipole amplitudes are suppressed compared to their real parts.
Therefore, in the analysis, only the real parts of the amplitudes were
kept. These amplitudes were parameterized as follows:
$E_{0+},S_{0+}\sim const$, $M_{1\pm}$,
$S_{1\pm}$, and $E_{1+}\sim p^*_{\pi}$.

In the fitting procedure,
the amplitudes $E_{0+}$, $S_{0+}$ and $M_{1\pm}$
were fitted without any restrictions. The relatively small
amplitudes $S_{1\pm}$ and $E_{1+}$ were fitted within ranges
found from the results of the analysis of the data \cite{ungaroprl,kijunprc2008}
using the UIM and DR in Ref.~\cite{aznauryan}.
It should be mentioned that the results for the $M_{1\pm}$
contributions obtained in our fit of the $\gamma^*p\rightarrow \pi^0 p$
cross sections near threshold are consistent with those of Ref.~\cite{aznauryan}
obtained in the analysis of significantly larger range over $W$. The overall
average $\chi^2$ per degree of freedom for the fit is approximately one. 

The obtained results for the structure functions are plotted
in Figs.~\ref{fig:fig14}-\ref{fig:fig16} as solid curves. 
It can be seen that the multipole
amplitudes $E_{0+}$, $S_{0+}$, $M_{1\pm}$,
$S_{1\pm}$, and $E_{1+}$ parametrized in the way discussed above
represent the data very well at all $W$.
The obtained results for $E_{0+}$ and $S_{0+}$
are presented in Fig.~\ref{fig:mult}. These multipoles have been
normalized to the dipole formula $G_D(Q^2) = \left(1 + \frac{Q^2}{.71}\right)^{-2}$.

Fig.~\ref{fig:g1g2} shows the extracted generalized form factors, $G_1$ and $G_2$, as
a function of $Q^2$. The error bars on the points include statistical and systematic 
uncertainties added in quadrature. The size of the estimated systematic uncertainties
is shown separately at the bottom of the plots, which assumes all systematic errors 
for all the data points to be entirely uncorrelated (10.8\%). 
The LET \cite{schkoch} predictions are shown as dash-dotted curves.

The plots also show LCSR predictions \cite{braunrev} 
as solid and dashed curves. Braun \textit{et al}.~have tried to minimize
the uncertainties in their LCSR based model calculations by including electromagnetic
form factor values known from experiment. These calculations are shown
as solid curves in the figure. The ``pure'' LCSR based models are calculations
where all the form factors are obtained entirely from theoretical calculations
and the uncertainties have not been minimized. These are shown as dashed curves 
in the figure. 
The difference between these two curves can essentially be treated as the overall
uncertainty in their predictions.


\section{Discussion}

The results for the $E_{0+}$ multipole and $G_1^{\pi^0p}$ are in good agreement 
with the LCSR predictions. The extracted $E_{0+}$ values deviate 
significantly from the LET predictions over the entire $Q^2$ range even though the 
extracted $G_1^{\pi^0p}$ values are not too far off from the LET predictions.
This is because the LET calculations for $E_{0+}$ only depend on $G_1^{\pi^0 p}$
(Eq.~(\ref{eqn:letmult})), whereas the LCSR calculations include 
contributions from both $G_1^{\pi^0 p}$ and $G_2^{\pi^0 p}$ (Eq.~(\ref{eqn:ffmultrelation1})).
The overall trends of increasing $E_{0+}$ and decreasing $G_1^{\pi^0 p}$ are similar 
to these two predictions, but the deviation of the extracted values for 
$G_1^{\pi^0 p}$ from the LET predictions becomes much more apparent at 
$Q^2 > 3$ GeV$^2$. 

One can observe a discrepancy of our results for the $S_{0+}$ multipole
and $G_2^{\pi^0p}$ from the LCSR predictions. The results are closer
to the LET predictions but are not entirely consistent for all $Q^2$. 

The uncertainty in the LCSR predictions for the $S_{0+}$ multipole 
and $G_2^{\pi^0p}$ is much bigger than for $E_{0+}$ and $G_1^{\pi^0p}$. 
In the chiral limit approximation, $m_\pi \to 0$, the Pauli form factor $F_2(Q^2)$,
which is the primary contributor to the calculations of 
$S_{0+}$ and $G_2^{\pi^0p}$, is not reproduced very well. Also, the LCSR 
calculations exist in leading order only and do not include next-to-leading order 
(NLO) corrections. The NLO corrections are expected to be large. 
Additionally, the LCSR predictions contain approximations and were not expected to 
have an accuracy of better than 20\% \cite{braunrev}. 

Furthermore, the LCSR predictions do not include effects from terms proportional 
to the pion mass. In the $Q^2$ region of this experiment, the predictions 
indicate a suppression of the $S_{0+}$ multipole \cite{braunrev} and this
multipole is very sensitive to corrections of all kinds, including the pion
mass corrections. In the LET predictions, some pion mass corrections have been 
included \cite{schkoch}. This may also explain the discrepancy between the 
predictions and the extracted results for $S_{0+}$ and $G_2^{\pi^0 p}$.  

Due to these 
theoretical uncertainties, the predictions of the magnitude of $S_{0+}$ and 
$G_2^{\pi^0p}/G_D$, and where they cross zero, differs for the two methods of
calculation. The experimental results indicate that this sign change for 
$G_2^{\pi^0p}/G_D$ occurs at $Q^2 > 4$ GeV$^2$ rather than at 
the LCSR prediction of around $2.2$ GeV$^2$ or $3.5$ GeV$^2$. 

The results of the structure functions, Figs.~\ref{fig:fig14}-\ref{fig:fig16}, 
indicate a significant contribution of the $p$-wave in the near threshold region as
indicated by the almost linear dependence of the $\sigma_T + \varepsilon\sigma_L$ as a 
function of $\cos\theta_\pi^*$. This contribution increases as one moves away from
threshold to higher $W$ (\emph{e.g.}, see Fig.~\ref{fig:fig16}). 
This is highly under-estimated in the overall LCSR 
predictions for the structure functions and cross section calculations. 
Their 
predictions are tuned to include mostly $s$-wave and very little $p$-wave 
contribution very close to threshold at high $Q^2$. This also explains the good
agreement of the extracted $E_{0+}$ and $G_1^{\pi^0p}$ 
to their predictions but the strong disagreement of the 
$S_{0+}$, $G_2^{\pi^0p}$, the cross sections and the structure functions. 

The extracted generalized form factors, $G_1^{\pi^0 p}$ and
$G_2^{\pi^0 p}$, show a faster fall off than the dipole form. This 
suggests a broadening of
the spatial distribution of the correlated pion-nucleon system. It suggests
that the correlated pion-nucleon system is broader than the bare nucleon itself
because the bare nucleon follows the dipole form factor.
 
The results for $G_1^{\pi^0 p}$ show similar trends to the 
previously extracted $G_1^{\pi^+ n}$ \cite{kijun2012}. In comparison, the former is 
about 30\% higher in magnitude while the overall behavior as a function
of $Q^2$ is similar. There are no results for $G_2^{\pi^+ n}$ for 
comparison. However, the generalized form factor results for the $\pi^0 p$
channel provide strong constraints on chiral aspects of the nucleon structure
and the validity of the LETs at high $Q^2$.


\begin{acknowledgments}
The authors thank Vladimir Braun for insightful discussions on the theoretical
aspects of this work. 
The authors also acknowledge the excellent efforts of the Jefferson Lab's 
Accelerator and
the Physics Divisions for making this experiment possible. This work was supported 
in part by the US National Science Foundation, the US Department of 
Energy, the Italian Istituto Nazionale di Fisica Nucleare, the French Centre
National de la Recherche Scientifique, the French Commissariat \`a l'\'Energie 
Atomique, the United Kingdom's Science and Technology Facilities Council, 
the Chilean Comisi\'on Nacional de Investigaci\'on Cient\'ifica y Tecnol\'ogica, 
the Scottish Universities Physics Alliance, and the National
Research Foundation of Korea. The Southeastern Universities Research Association
(SURA) operated the Thomas Jefferson National Accelerator Facility for the US
Department of Energy under Contract No.~DE-AC05-84ER40150.
\end{acknowledgments}

\bibliography{myrefs}

\end{document}